\documentclass{aa}  
\usepackage{graphicx}
\usepackage{txfonts}
\newcommand{\rms}{{\em rms}}

\newcommand{\period}{{\sc period98}}
\newcommand{\momf}{{\sc momf}}
\newcommand{\isis}{{\sc isis}}
\newcommand{\neda}{{\sc neda}}
\newcommand{\daogrow}{{\sc daogrow}}
\newcommand{\daomaster}{{\sc daomaster}}
\newcommand{\daomatch}{{\sc daomatch}}
\newcommand{\daophot}{{\sc daophot}}
\newcommand{\allstar}{{\sc allstar}}
\newcommand{\allframe}{{\sc allframe}}
\newcommand{\cf}{cf.}
\newcommand{\eg}{e.g.}

\newcommand{\Msun}{\hbox{$\thinspace {\rm M}_{\odot}$}}
\newcommand{\Rsun}{\hbox{$\thinspace {\rm R}_{\odot}$}}
\newcommand{\Lsun}{\hbox{$\thinspace {\rm L}_{\odot}$}}

\newcommand{\mhz}{$\mu$Hz}
\newcommand{\teff}{$T_{\rm eff}$}

\begin{document}
   \title{A search for solar-like oscillations in K~giants in the globular
   cluster M4\thanks{Based on observations with the Danish 1.54\,m on La~Silla at the European Southern Observatory in Chile,
the 1.5\,m at Cerro Tololo Inter-American Observatory in Chile, and the SSO\,40'' at Siding Spring Observatory in Australia.}}
   \author{S.\ Frandsen
          \inst{1}
	  \and
	  H.\ Bruntt
	  \inst{1,2}
          \and
          F.\ Grundahl
	  \inst{1,3}
	  \and
	  G.\ Kopacki
	  \inst{4}
	  \and
	  H.\ Kjeldsen
	  \inst{1,3}
	  \and
	  T.\ Arentoft
	  \inst{1,3}
	  \and
	  D.\ Stello
          \inst{2}
	  \and
	  T.~R.\ Bedding
          \inst{2}
	  \and
	  A.~P.\ Jacob
	  \inst{2}
          \and
          R.~L.\ Gilliland
          \inst{5}
	  \and
	  P.~D.\ Edmonds
	  \inst{6}
	  \and
          E.\ Michel
	  \inst{7}
	  \and
          J.\ Matthiesen
	  \inst{1}
          }
   \offprints{S.~Frandsen}
   \institute{Department of Physics and Astronomy, University of Aarhus,
              Ny Munkegade, Bygn.\ 520, DK-8000 Aarhus, Denmark
              \email{srf@phys.au.dk}
 	 \and
	      School of Physics A28, University of Sydney, 2006 NSW, Australia
	 \and
	      Danish AsteroSeismology Center (DASC), University of Aarhus, DK-8000 Aarhus, Denmark
         \and
              Institute of Astronomy, University of Wroc\l{}aw, Poland
         \and
	      Space Telescope Science Institute, Baltimore, USA
	 \and
	      Center for Astrophysics, Cambridge, MA, USA
	 \and
	      Observatoire de Paris, Meudon, France
             }
   \date{Received xxx 2007; accepted yyy 2007}
% \abstract{}{}{}{}{} 
% 5 {} token are mandatory 
  \abstract
   {To expand the range in the colour-magnitude diagram where asteroseismology 
   can be applied, we organized a photometry campaign to find evidence for 
   solar-like oscillations in giant stars in the globular cluster M4.}
   {The aim was to detect the comb-like $p$-mode structure characteristic 
   for solar-like oscillations in the amplitude spectra. 
   The two dozen main target stars are 
   in the region of the bump stars and have luminosities in the range 50--140\,$\Lsun$.}
   {We collected 6160 CCD frames and light curves for
   about 14\,000 stars were extracted.
   The frames consist of exposures in the Johnson $B$, $V$ and $R$ bands 
   and were obtained at three different telescopes. 
   Three different software packages were applied to obtain the lowest possible 
   photometric noise level. The resulting light curves have been
   analysed for signatures of oscillations using a variety of methods.}
% such as autocorrelation of the amplitude spectrum. 
% (*HB: Mention other methods, e.g.\ the comparison with simulations.)
   {We obtain high quality light curves for the
   K~giants, but no clear oscillation signal is detected.
   This is a surprise as the noise levels achieved in the
   amplitude spectra should permit oscillations to be seen at the
   levels predicted by extrapolating from stars at lower 
   luminosities. In particular, when we search for the
   signature of oscillations in a large number of stars we might
   expect to see common features in the power spectra, but even
   here we fall short of having clear evidence of oscillations.}
%   {\bf The RR~Lyr and the sdB star is not mentioned in the paper. Delete the next sentence?}
%   Very good light curves have been obtained for RR~Lyrae stars
%   and for a pulsating sdB star.}
   {High precision differential photometry is possible even
   in very crowded regions like the core of M4. Solar-like oscillations are
   probably present in K~giants, but the amplitudes are lower
   than classical scaling laws predict. 
   The reasons may be that the lifetime of the modes are short
   or the driving mechanism is relatively inefficient in giant stars.
   }
   \maketitle
%
%________________________________________________________________

\section{Introduction}

   Asteroseismology has made great progress in
   recent years due to improved observational techniques. 
   Extremely stable spectrographs have been designed with the aim 
   to detect exo-planets through Doppler shifts.
   These instruments have allowed for the unambiguous 
   detection of $p$-modes in about a dozen solar-type stars 
   (for a review see Kjeldsen \& Bedding \cite{review}). 
   Although most results have been 
   obtained for dwarf and subgiant stars, some giant stars have been
   shown to oscillate: $\xi$ Hya (G7III; Frandsen et al. \cite{xihya}), 
   $\epsilon$ Oph (G9.5III; De~Ridder et al. \cite{epsoph}), 
   and $\eta$ Ser (K0III; Barban et al.\ \cite{kgiants}).
   Unambiguous detection of $p$-modes has so far been restricted to field 
   stars in our immediate neighbourhood.    

%% ** HB-June-Rev.: the paragraph below rewritten slightly: see esp. last 2-3 sentences.  
%% ** HB-June-Rev.: checked the Dollinger paper: it is an ESO messenger paper. One a short
%% conf. proceedings from 2006 is available. They claim that the high RV in 36 stars 
%% in their sample is due to oscillations.  
   D\"ollinger et al. (\cite{messenger}) found that field stars 
   show radial velocity and photometric variability, 
   which seems to increase with increasing luminosity.
   This could either be due to a combination of activity and granulation or pulsations,
   either self-excited (Mira-like) or driven by convective motions (solar-like).
   Similar evidence for increased variability has been found
   for stars in the globular cluster 47 Tucanae (Edmonds and Gilliland \cite{tuc47}).

   Stellar clusters present a highly interesting prospect, 
   due to the large number of stars collected in a limited field of view 
   and additional constraints from cluster properties like 
   relative age, metallicity, and evolutionary stage.
   Gilliland et al. (\cite{m67}) were the first to make a search 
   for solar-like oscillations in the main sequence F-type stars 
   in the open cluster M67. 
   However, based on their multi-site campaign using
   4-m class telescopes they made no clear detection.
   Recently, another campaign was carried out on the same cluster
   (Stello et al.\ \cite{stello06,stello07}), this time
   concentrating on stars on the lower part of the giant
   branch with expected larger oscillation amplitudes.

   Knowing that $p$-modes are excited in giant stars, 
   we decided to study K~giants in the globular cluster M4. 
   The two main reasons are that M4 has a rich population of K~giants (see Fig.~\ref{fig:1a})
   and the amplitudes are expected to be substantially higher than for subgiants.
   The prospect is that if we can measure the gross properties of the oscillations,
   like amplitudes and the large splitting characteristic for solar-like oscillations,
   we could potentially study an ensemble of
   giant stars in different evolutionary stages. 

%% ** HB-June-Rev.: paragraph below slightly rephrased / reordered arguments.
   According to the empirical calibration by Kjeldsen \& Bedding (\cite{kandb}) 
   oscillation amplitudes scale as $L/(M\,T_{\rm eff}^2)$. For the K~giants
   in M4, taking the solar amplitude as 4.7\,ppm (parts per million),
   we expected amplitudes in the range 400--1000\,ppm.
   The noise level we expected to reach 
   in the amplitude spectra for the bright
   giant stars was $\simeq\,25$\,ppm at high frequencies. 
   This would allow us to detect any excess power and possibly
   the comb-like structure of the $p$-modes in the amplitude spectra 
   for the brightest and least crowded targets.
   The mode frequencies unfortunately decrease with luminosity
   and approach a range where noise caused by extinction and
   transparency changes is difficult to eliminate in ground-based data.

\section{The observations}

CCD frames were obtained with the Danish 1.54\,m telescope at La~Silla, 
the 1.5\,m telescope at Cerro Tololo Inter-American Observatory (CTIO), both in Chile,
and the 1\,m telescope at the Siding Spring Observatory (SSO) in Australia. 
The observations took place over a period of 
almost three months in order to get the necessary frequency resolution.
We started at La~Silla on April 13, 2001 and got the last data
on June 27, 2001.
We obtained useful data on 48 nights out of an allocation of 63 nights.
Some nights had overlap
between Australia and Chile which % will improve the window function
can be used to check the consistency of the photometric results.
%%{\bf DS:
%%  No, overlap of more than one site makes the window less good (more
%%  sidelobes) It is only when the data fill the 
%%gaps of each other that the window improves.}.
Table \ref{table:1} gives a list of the collected dataset.
In total we have 6160 CCD frames of M4 
(4014 from La~Silla, 863 from CTIO, and 1283 from SSO).

In order to identify the mode type, we decided
to observe in three filters $B$, $V$ and $R$. Amplitude ratios 
between different colours will indicate which type of
modes are present. To enhance the signal to noise ratio the data can still be 
combined using an approximate scaling (Eq.~\ref{eq:slut}) as phase
differences of $p$-modes between filters are small.  
About equal time was allocated to the three filters giving
more frames in $R$ due to the shorter exposure times.

%% {\bf HB: Mention here why we decided to use three filters, $BVR$, and
%% why we collected so many $R$ images. We should mention how we have combined
%% the light curves in the three filters. This was done by scaling the variation
%% -- how was the scaling determined? In the conclusion we should mention
%% if we recommend to use $BVR$ or just one filter in future campaigns. 
%% For M67 Stello et al. used almost only $V$ data (10\% of data points 
%% were taken in $B$).}

\begin{table}
\caption{Observing log for the M4 campaign in 2001.}             % title of Table
\label{table:1}      % is used to refer this table in the text
\centering                          % used for centering table
\begin{tabular}{lrrrll}        % centered columns (4 columns)
\hline\hline                 % inserts double horizontal lines
             & \multicolumn{3}{c}{Number of images}                                        & \multicolumn{2}{c}{Date of observation} \\
Observatory  & \multicolumn{1}{c}{$B$} & \multicolumn{1}{c}{$V$} & \multicolumn{1}{c}{$R$} & \multicolumn{1}{c}{Start} & \multicolumn{1}{c}{End} \\    
\hline                        % inserts single horizontal line
SSO          &   0 &  548 &  735  & May 1    & June 10 \\ 
CTIO         & 143 &  274 &  446  & June 1   & June 6  \\
La~Silla     & 847 & 1358 & 1809  & April 13 & June 30 \\      % inserting body of the table
\hline                                   %inserts single line
\end{tabular}
\end{table}

\begin{figure}
   \centering
   \includegraphics[width=80mm]{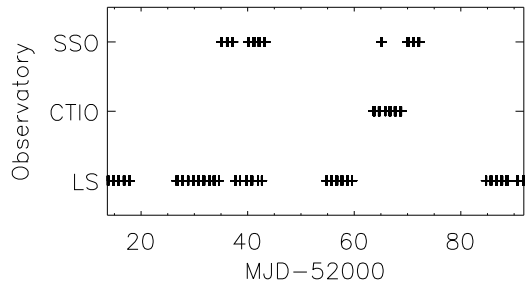}
   \caption{{Distribution of nights for the three observing sites. 
The time baseline is 78 days.} 
  }
              \label{fig:1}%
\end{figure}

The time distribution is represented in Fig. \ref{fig:1}, which includes
31 nights from La~Silla, 6 nights from CTIO, and 11 nights from SSO.

% {\bf HB: insert CMD here. 
% Perhaps Stetson has now made the standard $B-V$ photometry?}

\begin{figure*}
   \centering
   \includegraphics[]{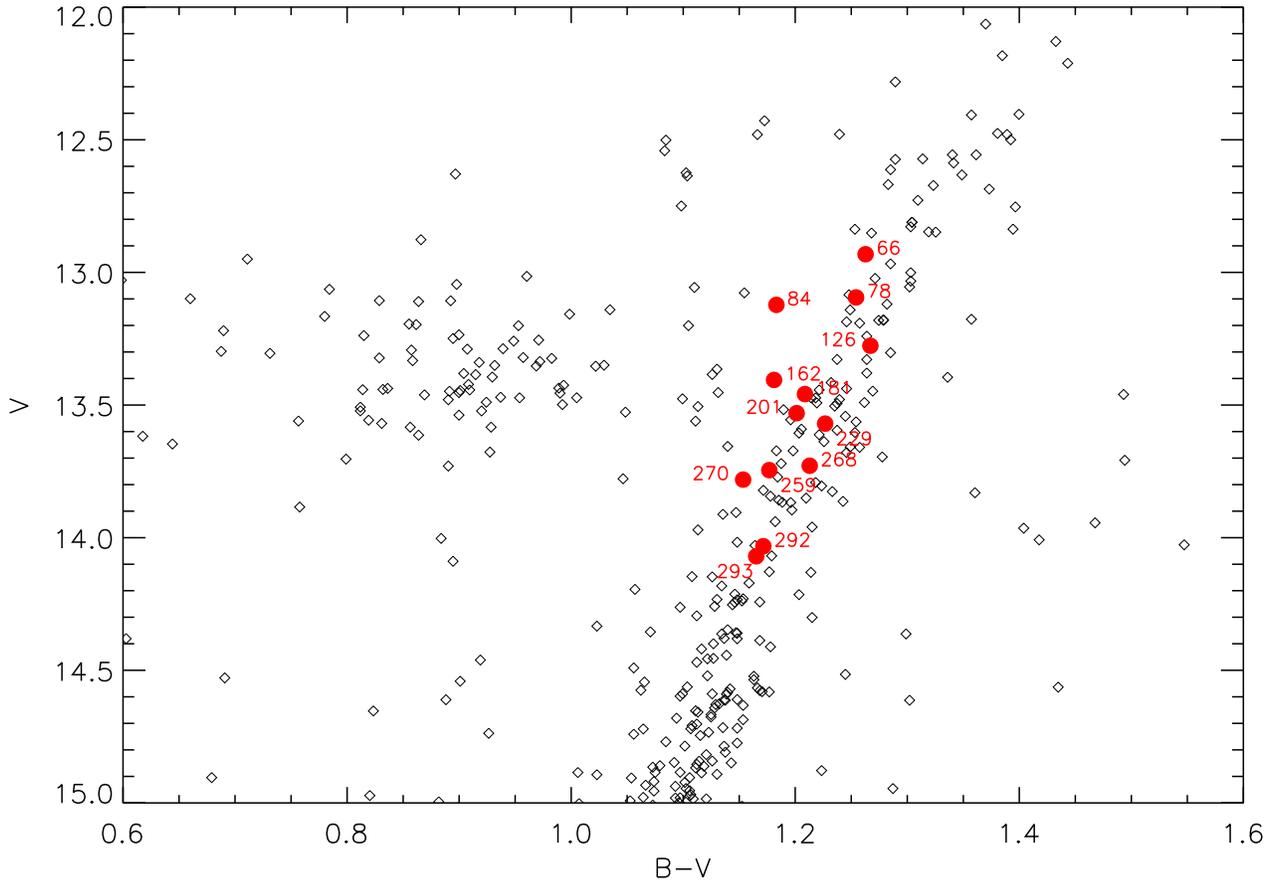}
   \caption{
   The colour-magnitude diagram for M4 showing the 
   bump stars and part of the horizontal branch. The colours and
   magnitudes have been calibrated using the database provided
   by Stetson (\cite{stetson07}). To show the range covered, some of the stars 
   selected for a detailed
   analysis are indicated and labeled with the ID -- see Table\,\ref{table:3}.}
              \label{fig:1a}%
\end{figure*}

The exposure times were adjusted to give the best result for stars
in the range $12.5 < V < 13.5$, which is the area of the RGB bump
in M4 shown in Fig.~\ref{fig:1a}. Typically this meant exposure times 
at La~Silla and CTIO of less than a minute in $V$ and $R$ and a few
minutes in $B$ depending on seeing conditions. 
Exposure times at SSO were 2--4 minutes in $V$ and 1--2 minutes in $R$.
On average this led to a mean duty cycle around 50\%, since the readout
time was 1--2 minutes. 

% HB-June-Rev.: minor detail, not used in the paper:
%At the Danish 1.54\,m telescope we read out the CCD using two
%amplifiers to reduce the readout time from 80\,s to 40\,s and thus
%improve the duty cycle. 

% This comment removed by HB: it's a very important point for people interested 
% in improving the cadence of time series photometry.
% {\bf DS: The next paragraph is non-important details that make the
%  reader fall asleep. You already made a good summary of exposure and read
%  out time and duty cycle in previous paragraph. This one does not add
%  clearification to that, only the opposite. In other words you dont use
%  that information later on and hence is redundant for the paper.}

%% The combination of the two half images
%% introduced some additional work during the calibration and
%% reduction.

%%{\bf DS: This paragraph is also non-important details that make the
%%  reader fall asleep...why do you tell me you used CCD number 2...who
%%  cares. It even requires that the reader knows that you used
%%  a mosaic CCD...but just delete it all.}
%% At SSO we read out four CCDs in parallel, but only the data from
%% one of them (\#2) have been used\, since stars in the other frames are not present
%% on the frames from La~Silla and CTIO.

\section{CCD photometry techniques\label{sec:ccd}}

The images were first calibrated by performing traditional bias
subtraction and flat fielding. In order to avoid nightly
offsets, calibration frames were constructed for periods of several
days, typically a week, and the same calibration frames were used for
all images in that period. This point is quite important since 
the timescale of variability for the giant targets is in the
range 9--31 hours (cf. Sect.~5.1).

CCD non-linearity was considered, and in the case of the La~Silla data
the non-linearity was corrected by the technique described by Stello et
al. (\cite{stello06}). The other cameras did not provide the
same option for a correction and were assumed to be linear.
We have not seen any evidence for non-linearity, and as we
were able to find comparison stars with similar magnitudes in
all cases, the influence of non-linearity should be negligible.

\begin{figure*}
   \centering
   \includegraphics[width=\textwidth]{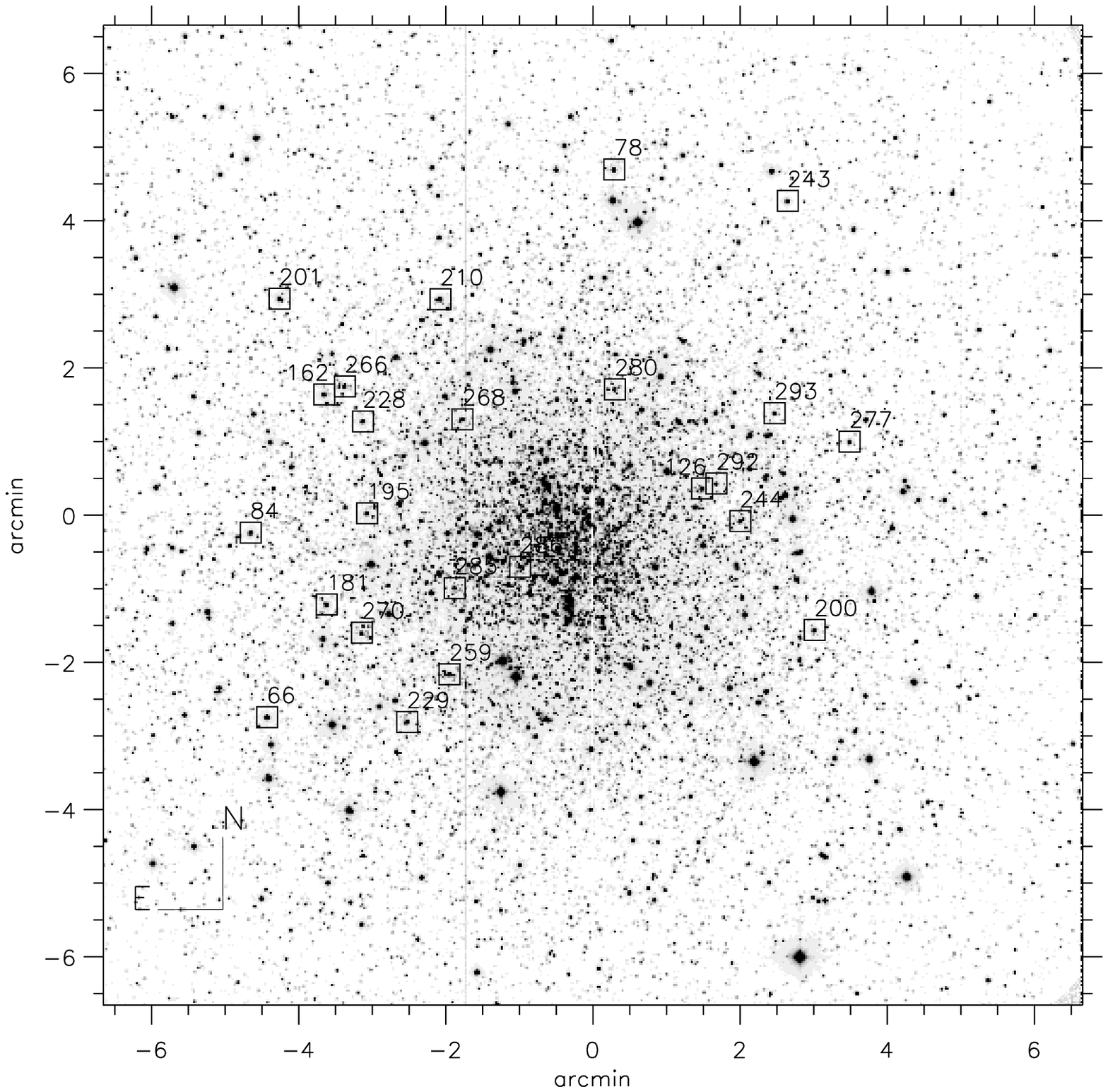}
   \caption{  % {\bf DS: Put axis labels on. Look at ex fig 3 (MNRAS 373 1141)}
 	The field of M4 observed with the Danish 1.54\,m telescope at La~Silla. 
        The two dozen K~giants we analysed in detail are marked by squares.
        The center of the field
 	is at $\alpha_{2000} = 16^h23^m33^s.7,\,\delta_{2000} = -26^{\circ}31'05".7$.
              \label{fig:2}}
    \end{figure*}

In Fig. \ref{fig:2} we present the calibrated CCD image, which has
been used as the reference image for the La~Silla observations of M4.
The two dozen K~giant stars we selected for
a detailed analysis are marked by boxes.

% {\bf HB: the SdB star is marked in Fig.~3 and mentioned in the caption but
% it is not mentioned anywhere else in the text.
% I strongly suggest removing any mention of the star and also no plot the diamond in Fig.~3.}

%%%
%%% HB-urgent: The ID numbers of 162 and 126 should 
%%% be moved slightly (overlapping with other ID numbers!)
%%%
%%% SF - done!!
%
% Caption text removed by HB:
% The single diamond is the subdwarf (O'Toole\,\cite{sdb}).
%        {\bf DS:what do you mean by THE subdwarf}. The center of the field
% 	is at $\alpha_{2000} = 16^h23^m33^s.7,\,\delta_{2000} = -26^{\circ}31'05".7$.
% {\bf HB: Give approximate RA/DEC and indicate directions of north/east? I think Frank has the info}
% {\bf HB: The subdwarf is not mentioned anywhere else. Delete it or give a reference to the paper?}

In order to achieve the highest possible photometric precision we decided
to use three different reduction programs, which have all been
successful in the past in producing high quality results: 
\isis, \daophot\ and \momf. %%% . 
In the most crowded areas we expected \isis\ (Alard \& Lupton \cite{isis98}; Alard \cite{isis00}) 
to give the best results.
In the less crowded areas we expected that \daophot\ (Stetson \cite{stetson})
would do best.  Finally, we know that \momf\ (Kjeldsen \& Frandsen \cite{momf}) 
performs very well in fields with only mild crowding. All frames were reduced with
\isis\ and \daophot\ and a subset of the $V$ frames from La~Silla were reduced with \momf. 
Each of the three methods will be described in Sects.~\ref{sec:isis}--\ref{sec:momf}
and we compare the results in Sect.~\ref{sec:comp}.

\subsection{The image subtraction method (\isis)\label{sec:isis}}

The main advantage of the difference-image technique is the 
ability to extract high precision photometry
in the crowded regions near the core of M4. 
In addition, any variations due to airmass and 
transparency changes are removed to first order 
as a part of the image subtraction.

The image subtraction method \isis\ (Alard \&\ Lupton \cite{isis98}; Alard
\cite{isis00}) was used by two members of the team independently (\isis1\ and \isis2).
For each filter and each observing site we selected the 
images with the best seeing to make the reference images. 
While \isis2\ only used a single reference image for each site and filter,
\isis1\ used subsets of data from about one week.

For each observed image, \isis\ computed a kernel 
which describes the variations of the PSF across the 
image relative to the reference image. 
\isis\ then convolves the reference image
with the kernel and subtracts this from the observed image. 
The resulting difference image will contain the
signal that is intrinsically different from the
reference image, \eg\ cosmic ray hits, hot pixels, 
and variable stars.

In the first approach (\isis1) only La~Silla and CTIO data were included (Bruntt \cite{bruntt_phd}). 
They were divided into five subsets and reduced separately. The reason for
this was large differences in the positional angle of the CCD camera
between different periods. We had problems using the photometry package that is part of \isis.
Instead, we modified the aperture photometry routine in \daophot\ in order to
be able to run it on the subtracted images, in particular allowing negative flux values.
The necessary modifications were described in detail by Bruntt et al.\ (\cite{bruntt6791}).
\isis\ needs a reference flux for each star, as only the change in the flux
is calculated. This was taken from the \daophot\ reduction of the reference images.

In the second reduction approach (\isis2) we constructed 
only one reference frame (for every observing site and passband).
The original \isis\ software does not work correctly with frames that
are significantly rotated with respect to the reference frame. The 
problem resides in the interpolation process. We used our own
application, which performs this task correctly even for large
rotation angles but requires much more computation time.
For all stars detected in the reference frames 
(of all three observing sites) we derived differential 
fluxes from our CCD frames following standard procedure
of \isis\ reductions (for details see Kopacki \cite{kop2000}). 
We performed both the PSF fitting and aperture photometry
on the difference frames. To transform differential 
fluxes into magnitudes we used
total fluxes measured in the reference frames using
aperture photometry tasks \neda$+$\daogrow\ under \daophot. 
Finally, all the data were merged into a uniform magnitude system by
applying magnitude offsets determined from a carefully chosen set
of bright unsaturated stars common to all three observing fields.
In this way, the CTIO and SSO data were transformed into the magnitude 
scale of the La~Silla data.

\begin{figure}
   \centering
   \includegraphics[width=80mm]{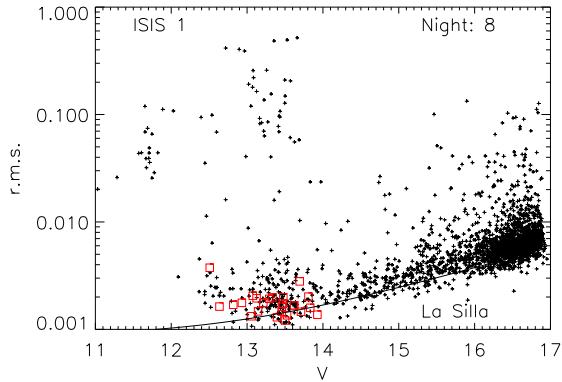}
   \caption{
% {\bf DS: Be consistent with labeling. Use ex V throughout not V
%         magn. and use rms or $rms$ or r.m.s. as in text...and no titles did
%        I say that before?}{\bf HB: I suggest using red colour for the box symbols: they are a bit hard to see.}
        The \rms\ scatter around the mean for a good
        night with the Danish 1.54\,m telescope for the \isis1 reduction. 
        The (red) squares give the noise
        level for the final time series for the selected K~giants,
        where we have looked for solar-like oscillations.
        The line indicates the estimated noise from scintillation and
        photon statistics.
        The excess of stars with high noise near $V=13.3\pm0.3$
        are the RR~Lyrae stars. Saturation sets in at $V \sim 12.5$.}
              \label{fig:3}%
    \end{figure}

\begin{figure}
   \centering
   \includegraphics[width=80mm]{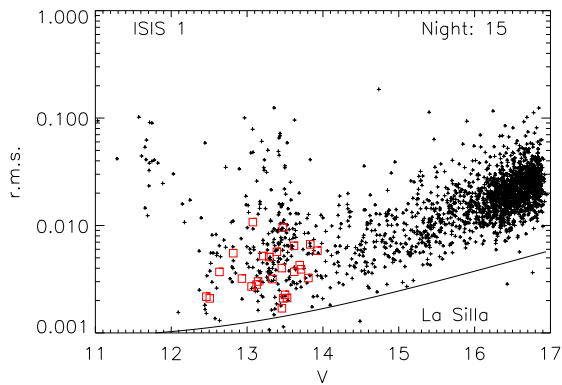}
   \caption{
%  {\bf DS: Consistent labeling!!!}
%  {\bf HB: I suggest using red colour for the box symbols: they are a bit hard to see.}
        Same as Fig.~\ref{fig:3}, but for a bad night. The night is characterized
        by very large sky background contribution due to moon light.}
              \label{fig:4}%
    \end{figure}

\subsection{\daophot/\allstar/\allframe\ reduction\label{sec:daophot}}

For the photometry we used the suite of photometry programs developed by
Stetson (\cite{stetson}, \cite{stetson2}, \cite{stetson3}): 
\daophot, \allstar\ and \allframe.
The general use of these is well described in manuals and the literature. For 
our applications we adopted a slightly modified procedure, which consisted
in the following.                  %%% HB-urgent: Dennis thought the next few sentences were unclear: Frank?
Firstly, a few of the deepest and lowest-seeing frames were
used to produce a master list of stars, which was obtained by several 
iterations through \daophot, \allstar\ and \allframe. Next, all frames were run
through \daophot/\allstar\ once. With the preliminary photometry in hand,
positional transformations between each frame and our reference frame
were derived using \daomatch\ and \daomaster\ (kindly provided by P.\ Stetson).

From the frames used to derive the master star list, we created two lists of stars
for generating the PSF for the individual images. The 
first list contained $\sim$10 stars that were well isolated and these were used
to generate the first version of the PSF for each frame. Next the 
second list, containing $\sim$200 stars covering the observed field, was 
used to generate the final PSF. We used this PSF when running \allframe\ on all
the images to produce the final photometry. 
We made a cross id of all stars in the photometry files 
and the 7,000 brightest stars were selected for further analysis.

% HB+DS-june: minor detail, only important for Frank.
% Due to the large number of 
% images and the extensive star list, we split the \allframe\ reductions into
% jobs containing 200 frames (but identical star lists). This allowed us
% to submit jobs to several computing nodes to help reduce the computing time. 

\subsection{\momf\ reduction\label{sec:momf}}

As an independent test we used the \momf\ package 
(Kjeldsen \& Frandsen \cite{momf})
on the La~Silla images to derive time series in the $V$ band. The
reduction had to be done in a slightly different way than normal, as
the La~Silla images were slightly rotated relative to each other. \momf\ can
only handle shifts. Thus we had to introduce an additional coordinate
transformation changing the $(x,y)$ coordinates for each image
before doing photometry. For stars outside the very crowded regions,
results were comparable to the other techniques. 
We conclude that we reach similar results for non-crowded regions by
all of the techniques.

\subsection{Comparison of the photometry\label{sec:comp}}

%% We have collected the photometry for each of the three reduction techniques. 

Our database contains photometry from the three different reduction methods 
and comprises light curves of 13\,611 stars with up to 6\,160 data points.
We also store information about each frame such as
seeing, airmass, background level, the mid-time of each observation 
(Julian date), and position on the reference frame of each star.
The data can be accessed at the web page {\it http://astro.phys.au.dk/$\sim$srf/M4/}.

The \rms\ scatter in the light curves is plotted versus magnitude 
for a relatively good and bad night in Fig.~\ref{fig:3} and \ref{fig:4}, respectively.
These results are based on the \isis1\
photometry in the $V$-band.
For the bad night, the noise is considerably larger than the good
night for all stars.
The higher noise is due to high sky background
as a result of the close proximity of the Moon to M4.
The lower envelope of stars for the good night 
is close to the expected precision based on the flux level (indicated by the solid line).
We find a number of stars with \rms\ scatter as low as 1.5\,mmag. 
For the K~giants the level is typically about 2\,mmag per data point.
The {\it bump} stars are found around $V=13.5$,
where we obtain the lowest noise levels. 
% Saturation of the CCD sets in at $V=12.5$. 
% The stars between 13.0 and 13.5 with very high \rms\ scatter are the RR~Lyrae stars.

\begin{figure}
   \centering
   \includegraphics[width=80mm]{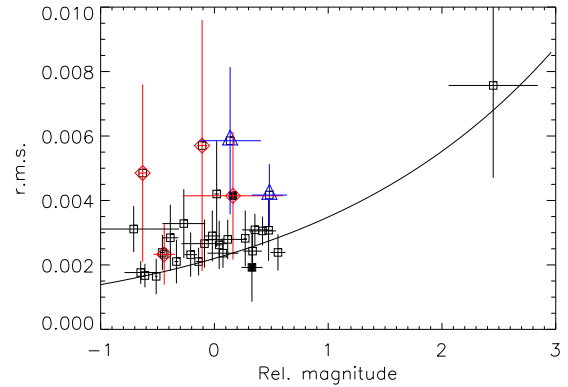}
   \caption{ %%% HB-urgent: the units on the y-axis should be multiplied by 1000 --
             %%%            In this way the units are the same as in Fig. 4,5,7,8 & 9 !!
             %%%            Fig. 6 is the only one with this unit (mmag) !!
             %%%
             %%% SF - done!!!!
      The mean \rms\ scatter per data point per night is plotted against
      the nightly mean flux level in the exposures represented by a
      corresponding magnitude difference between each night and a reference 
      night. The filled symbols are two nights with very few 
      observations. The over-plotted diamonds indicate nights with a
      high sky level and the triangles nights with very good seeing.
      The \rms\ uncertainty is marked for each point.
              \label{fig:5}}
    \end{figure}

\begin{figure*}
   \centering
   \includegraphics[width=80mm]{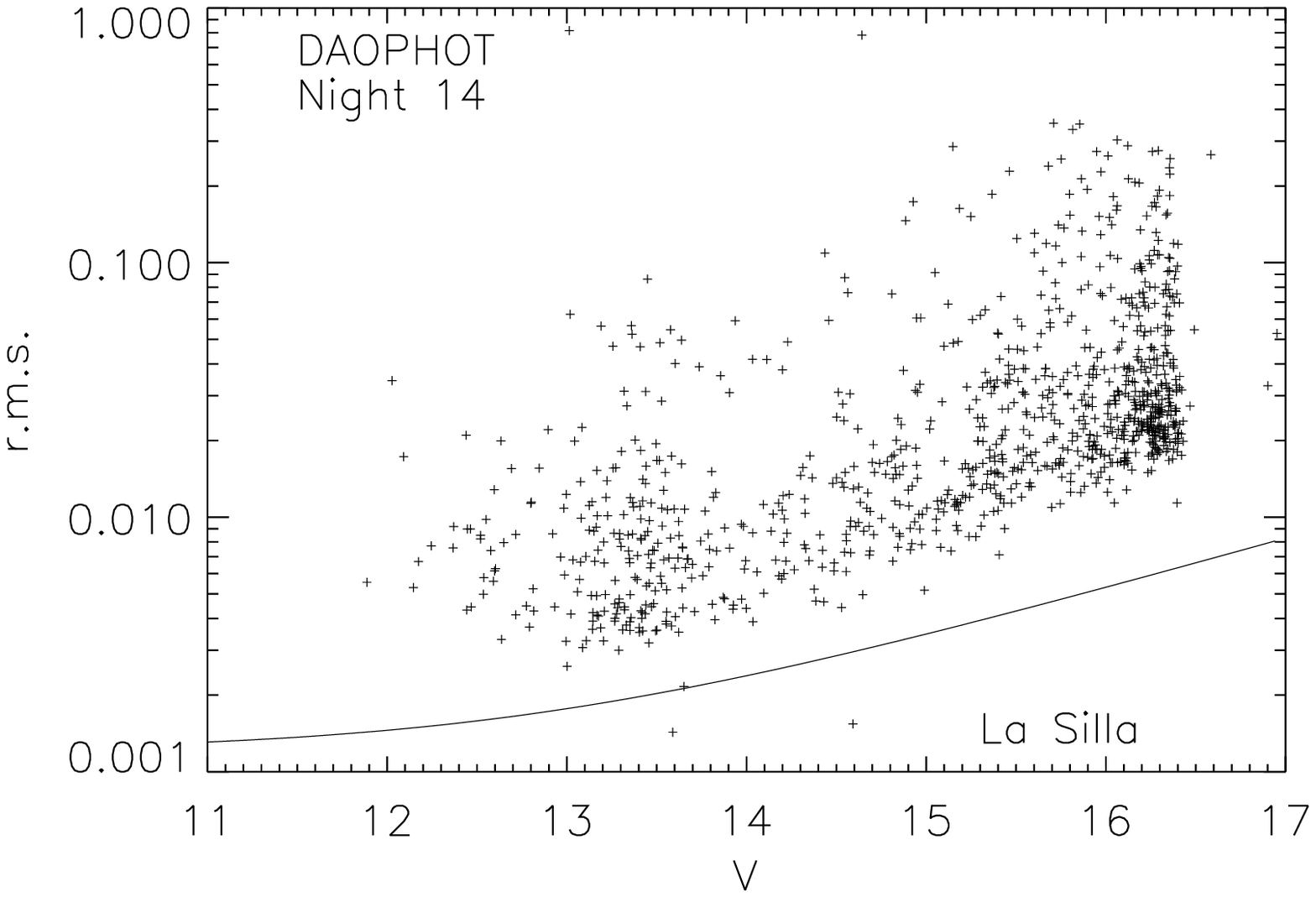}
   \includegraphics[width=80mm]{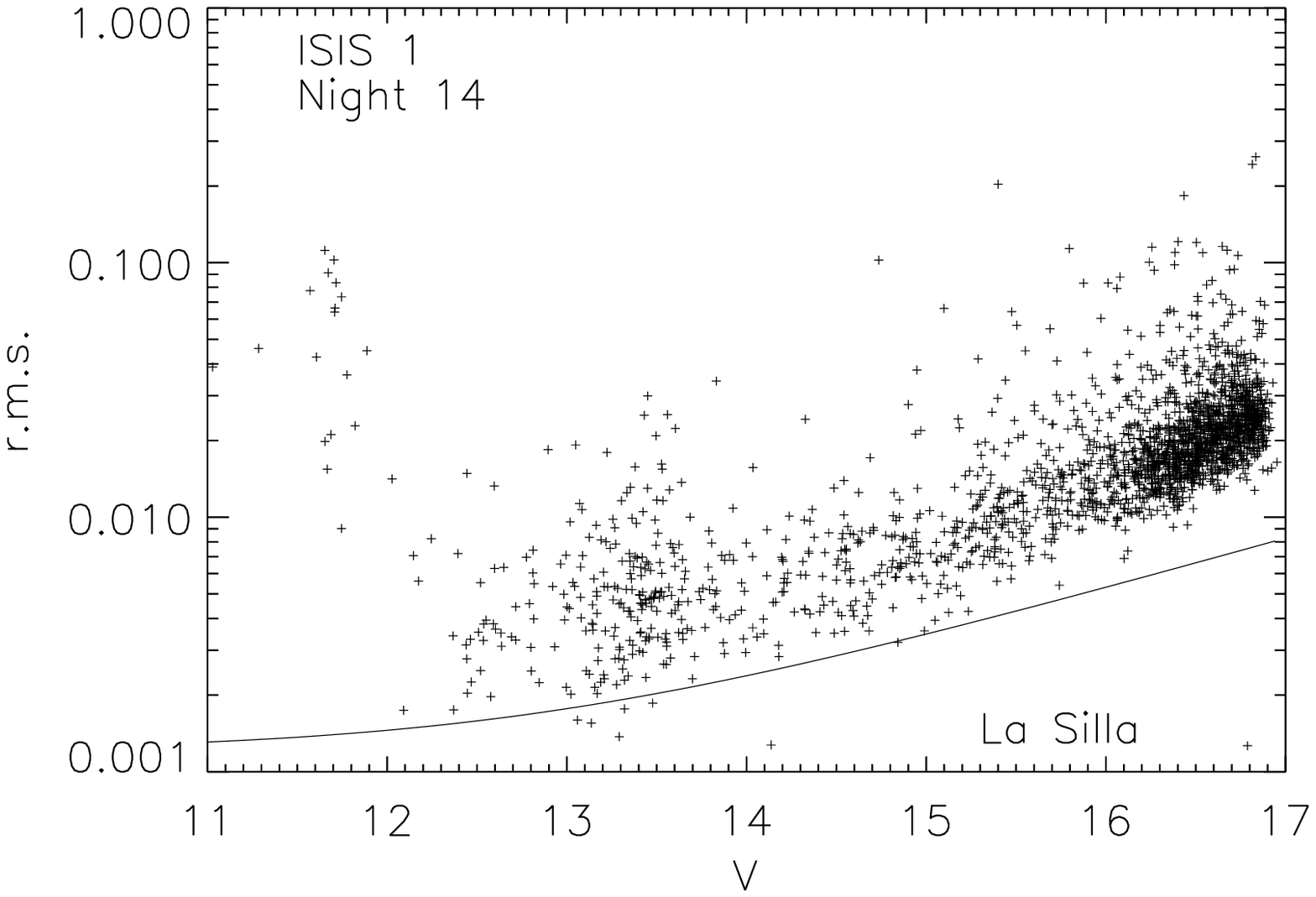}
   \caption{The mean scatter per data point per night for night 14 for 
            the \daophot\ (left panel) and \isis1 reduction (right panel).
            The line gives the noise level for a good night, see Fig.~\ref{fig:3}.
      }
              \label{fig:6}%
    \end{figure*}

%% \begin{figure}
%%    \centering
%%    \includegraphics[width=80mm]{scatter14_ls_dao.eps}
%%    \caption{Same data as in Fig.~\ref{fig:6}, but reduced with \daophot.
%%       The lower envelope of the data is not far from the results in
%%       Fig.~\ref{fig:6}, but the number of points with excess noise is
%%       clearly larger.}
%%               \label{fig:7}%
%%     \end{figure}

%% DS-june: Dennis thinks the next paragraph + figure could be removed:
As shown in Fig.~\ref{fig:4}, some nights do not give good results, with noise
levels considerably higher than the majority of the data. Naturally, the noise per data point
depends on the level of exposure per frame. This will vary as we had to adjust
the exposure time to avoid saturation. In good seeing we obtained fewer photons
per frame as the stellar image size is smaller and the maximum exposure level is kept constant
to avoid saturation.
Also, when cirrus clouds were present we kept the exposures times a bit conservative
in order to avoid occasional over-exposed images from holes in the clouds.

We have investigated in which conditions the photometry was 
worse than expected from photon noise alone. 
To do this we considered the \isis1\ photometry of stars from La~Silla.
We selected an ensemble 
of 19 bright stars and calculated their average scatter per night.
The average flux for the ensemble was also derived and converted to a magnitude. 
One good night was chosen as the
reference night and the magnitude for
that night was subtracted from the magnitudes from all other nights.

The measured \rms\ scatter is plotted against the magnitude for all the nights in Fig.~\ref{fig:5}.
If only photon noise were present the scatter 
should follow the solid line in Fig.~\ref{fig:5}, ie.\ following the
relation: $5\,\log_{10} \sigma = m-m_{\rm ref}$.

Most of the nights follow this law, but there are exceptions. Nights with a
high sky background clearly stand out as less valuable for the purpose
of achieving high photometric precision (diamonds in Fig.~\ref{fig:5}). 
A second problem arises, when the
seeing is very good since under-sampling sets in (triangles in Fig.~\ref{fig:5}).

%%% ({\bf give pixel scale here or in a Table with 
%%% properties of the telescope / CCD detector at the three sites?}).
%%% 
%%% 
%%% \begin{table}
%%% \caption{{\bf HB: The scale information is not used anywhere in the paper. 
%%% No reference is given to the Table. Perhaps delete it?}
%%% Image scale of the CCD cameras for the three sites.}             % title of Table
%%% \label{table:2}      % is used to refer this table in the text
%%% \centering                          % used for centering table
%%% \begin{tabular}{lr}        % centered columns (4 columns)
%%% \hline\hline                 % inserts double horizontal lines
%%% Observatory  & pixel scale in "\\
%%% \hline                        % inserts single horizontal line
%%% La~Silla     & 0.39 \\
%%% CTIO         &  0.43 \\
%%% SSO          &  0.38 \\
%%% \hline                                   %inserts single line
%%% \end{tabular}
%%% \end{table}
%%% Finally, when the flux level changes a lot during the night, we tend
%%% to get less good results. These are typically nights with cirrus and
%%% large transparency variations. Together, these three reasons are able
%%% to explain all nights with excess noise as shown in Fig.~\ref{fig:5}.

An important issue is whether any of the three reduction techniques are superior.
In general we find that the results are quite similar for the image subtraction, 
PSF fitting, and aperture photometry techniques.
However, \momf\ is clearly not suited for crowded fields and only for a
small fraction of stars does it perform as well as \isis\ or \daophot.
The quality of the \daophot\ photometry is comparable to the \isis\ results,
although \isis\ is superior in the most crowded areas. 
We also find the best results with \isis\ for faint stars, especially on nights with bad seeing.
This is shown in Fig.~\ref{fig:6} where
we compare the \rms\ scatter in data from one bad night from
La~Silla using \daophot\ (left panel) and \isis\ (right panel).
The \rms\ noise of the K~giant stars is 40\% lower for the \isis\ reduction.
%% $4.8$ to $6.7$ mmag for the K~giant targets.

It is possible that with a more careful choice of 
PSF stars and other parameters involved in the \daophot\ reductions, it
may be possible to improve these results. The optimum choice of parameters
for good observing conditions might not be the best for frames obtained under
non-photometric conditions. 
% But it is extremely time consuming to treat 
% the data in an optimum way by dividing the data in small subsets.
% This also introduces non-trivial problems when merging all subsets.

\begin{figure*}
   \centering
   \includegraphics[width=80mm]{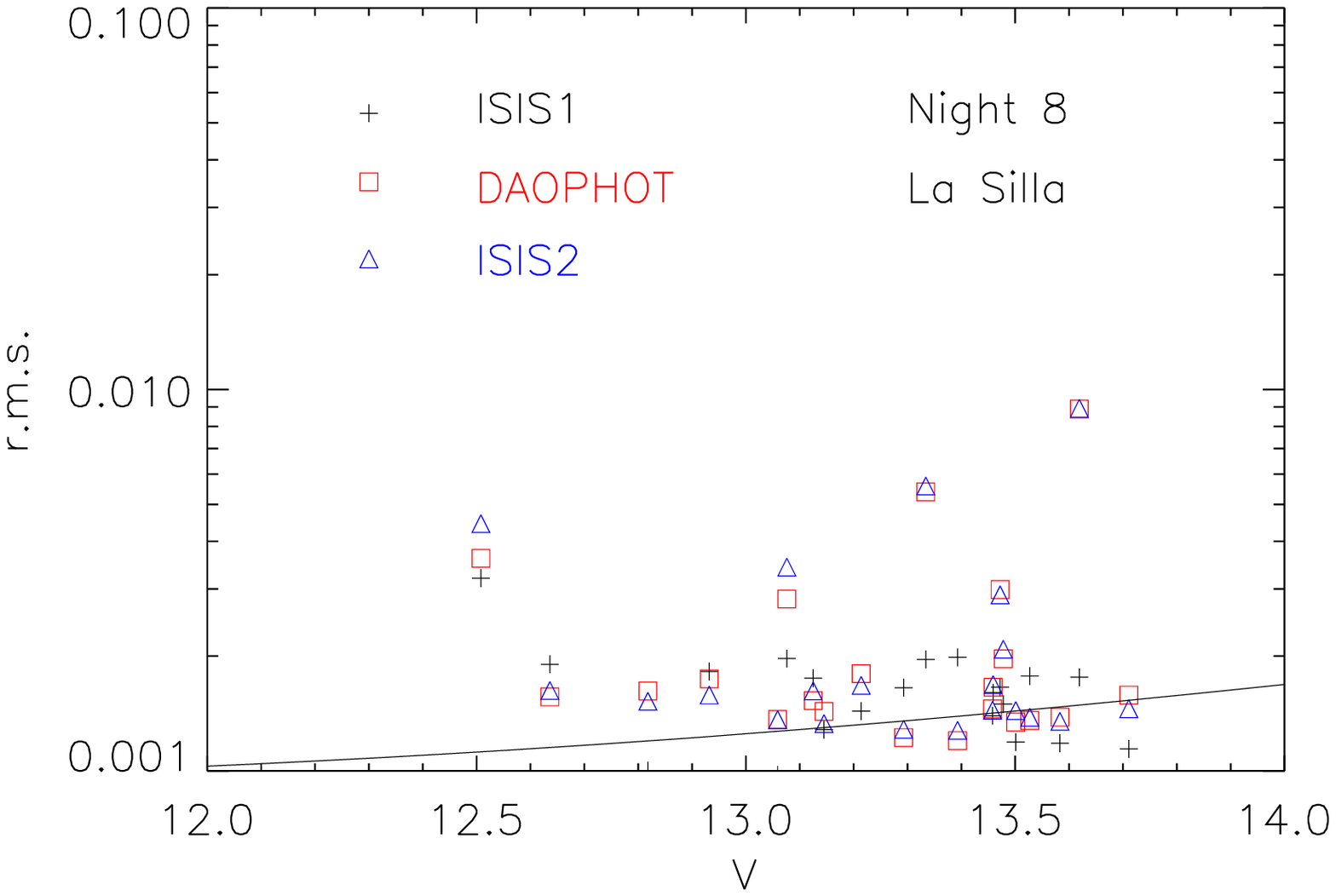}
    \includegraphics[width=80mm]{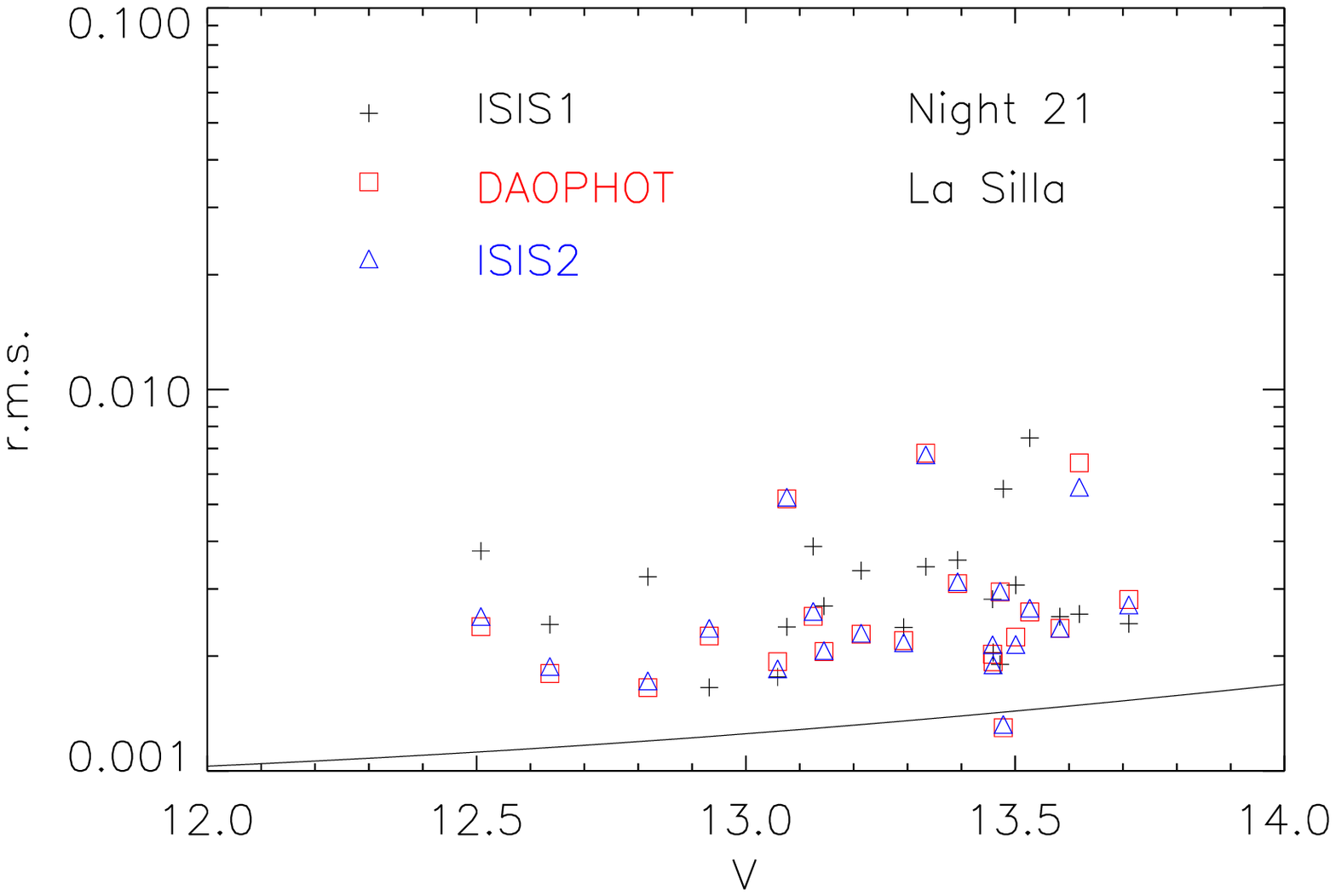}
   \caption{The \rms\ noise level in K~giant stars for La~Silla data from two different nights 
is shown in the left and right panel. 
We compare the noise level from the three reduction methods. 
It is similar for most stars, but in a few cases
the \daophot\ and \isis2 noise levels are higher than for \isis1.}
              \label{fig:7}%
    \end{figure*}

A final illustration of the differences seen between the different
algorithms is presented in Fig.~\ref{fig:7}. 
Here the \rms\ scatter 
is presented for three different reductions: \isis1, \isis2 and \daophot. 
Each point is for a particular star and the symbols define
which reduction technique was being used. It is evident, that none of
the techniques produce the best results in all cases. It seems
that the image subtraction method \isis1 has been most successful on average.

%% \begin{figure}
%%    \centering
%%    \includegraphics[width=80mm]{comp21_ls.eps}
%%    \caption{Same data as in Fig.~\ref{fig:8}, but for a different night.
%%        On this night \isis1 is not doing as well and the noise per data point
%%        is generally higher.}
%%               \label{fig:9}%
%%     \end{figure}

\begin{figure}
   \centering
   \includegraphics[width=80mm]{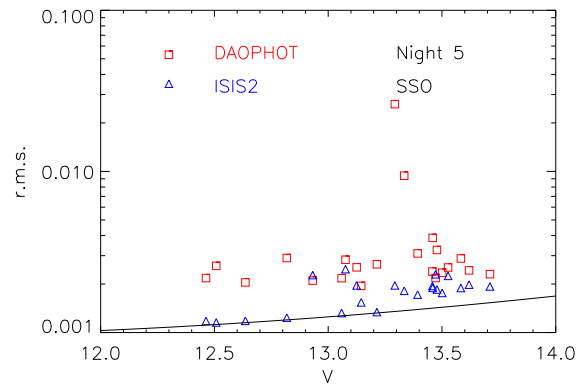}
   \caption{The \rms\ noise levels in data from a good night at SSO.
       \daophot\ is doing significantly worse than \isis2.
       Notice the increased noise at faint magnitudes
       relative to the photon noise estimate (solid curve).
       }
              \label{fig:10}%
    \end{figure}

The time series from each observatory show different
characteristics due to different observing conditions and
different telescope and detector setups. 
At the SSO the seeing was considerably
worse leading to more extended stellar images. This leads to increased
noise as a result of the higher influence of the background level,
which is worse for fainter stars. This effect is shown in Fig.~\ref{fig:10}.
The CTIO data are heavily influenced by the presence of the
Moon bing close to the target field, particularly in the middle
of the run, and this makes the data of rather poor quality.
The results can be characterized by the \rms\ noise per
data point for a non-crowded K~giants for the three sites on a good night:
$1.6$\,mmag for La~Silla, $3.0$\,mmag for CTIO and $2.1$\,mmag for SSO.

%% The CTIO value reflects the bad observing conditions. %% This much is clear.

\section{Light curve preparation}

% {\bf HB: Need to introduce the content of the following subsection. Why did we
% use ensemble photometry? Reading the preceding Section one might thing that the light curves
% are ready for analysis.}
The outcome of the previous section is either absolute or differential
magnitudes. In both cases it is necessary to make
differential photometry by comparing each target star
with a set of reference stars.
The effects of extinction, instrumental drifts and other sources
of non-stellar noise need to be 
minimized for us to be able to study and stellar signal at frequencies
below $30$\,\mhz\ corresponding to periods longer than 9 hours.
The approach is described in the following Sections.

\subsection{Ensemble photometry\label{sec:ensemble}}

An ensemble of reference stars is
used to derive the relative magnitude for all stars. This
should take out transparency and extinction variations. 
In M4 there is a wide range of possibilities for choosing reference stars. 
However, the results depend strongly on the implementation
of the ensemble average performed. 

% HB: removed these four lines: the next paragraph is much more specific!
% The simplest choice is a small set of stars ($\sim$10),
% which is used for all stars on all frames. We have attempted to make
% the selection more sophisticated by choosing different reference sets
% for different target stars on different nights. 

The algorithm we employed was the following:
for a given star, a number of reference stars were selected
which were required to have similar $V$ magnitude and $B-V$ colour and 
be within a certain distance (eg. half the size of the CCD) on the CCD frame. 
We calculated the magnitude difference of the target star and each reference star 
for a whole observing block ($t \ge 7$\,d) 
and subtracted the median value producing a time series $\Delta m_i(t)$,
where the index $i$ indicate the reference star.

We now do an analysis night by night.
For each night the \rms\ noise $\sigma_{i,\rm rms}$ of each $\Delta m_i(t)$ time series 
is calculated and a weight is assigned using
\begin{equation}
w_i = 1/\sigma_{i,\rm rms}^2.
\end{equation}
The 10 stars with highest weights are used to
calculate the final light curve for the night as the weighted sum
\begin{equation}
\Delta m(t) = \sum_{i=1}^{10} w_i \Delta m_i(t)/\sum_{i=1}^{10} w_i.
\end{equation}
The set of reference stars will vary from night to night. 
The method does not imply any high-pass filtering at a timescale of
one day, as the median is subtracted before splitting the time series
into single nights.
% {\bf DS: Remove following sentence...not relevant}
% Stars for which the photometry is unsuccessful on some nights, 
% are eliminated from the reference sets on these nights.
The technique gives slightly lower (10--20\%) white noise levels in the final
light curves compared with the simpler choice of a common
set of reference stars extending over all nights of an observing
period. The final light curves are obtained by selecting, for
each observing site, the result with lowest noise among the
different methods applied (\daophot, \isis1 and \isis2).

\subsection{Decorrelation}

%% HB rephrased this short section slightly, trying to make both SRF + DS happy:
%% See M4giants05.tex for DS' detailed comments.

Instrumental drifts and reduction noise may lead to
correlation of the measured magnitudes with parameters 
like the colour of the star, position on the CCD and airmass. 
We found no clear correlations between the relative magnitudes
and parameters like airmass, sky background or position on the CCD.
As a consequence no type of decorrelation was performed.

\subsection{Combining $B$,$V$ and $R$}

The results from the procedures described in Sect.~\ref{sec:ensemble}
are a set of light curves from the three sites in three filters
$B$, $V$ and $R$. 
In our search for solar-like oscillations in the K~giants we assume
that phase shifts between observations in different filters are small 
%%\bf DS: maybe give ref to 
(Jim\'enez et al.\ \cite{phases}). 
We also assume that the oscillation amplitudes scale
with inverse wavelength as $A \propto \lambda^{-1}$ (Eq. \ref{eq:slut}) and
we therefore
scaled the $B$ and $R$ time series relative to the
$V$ filter: scaling factors were 1.222 and 0.846.
The weights, described in the next subsection, were also scaled 
with the inverse factor following Eq.\,\ref{eq_w2}.

%%% HB-urgent: In the Table below the ID2 numbers are missing. Since you have the B-V, V magnitude 
%%% from Stetson, you should also have the ID2 values?
%%% I removed the ``.'' following all Teffs and dL/L ...
%%%
\begin{table*}
\caption{The 24 K~giants selected for detailed analysis. ID2 is from Stetson (\cite{stetson07})
if present in the catalog (NA means not available)
and $V$ and $B-V$ are calibrated using this database.
\teff\ is found using the $B-V$ calibration of Alonso et al.~(\cite{alonso}) while
the luminosity and radius were determined using the parameters for the M4 cluster (see text for details).
In the last four columns we list the predicted values for the peak amplitude in ppm when assuming $\delta L/L \propto (L/M)^{0.7}$,
$\nu_{\rm max}$ is the expected peak frequency for $p$-modes, 
$\Delta\nu_{0}$ is the large separation, 
and $\omega_c$ is the acoustic cutoff frequency (frequencies are all in $\mu$Hz and calculated from Eqs.~\ref{eq:start}--\ref{eq:slut}).
}
\label{table:3}                    % is used to refer this table in the text
\centering                         % used for centering table
\begin{tabular}{rrccrr|rrr|rrrr}   
\hline                             % insert a single horizontal line
 ID & ID2 & $\alpha_{2000}$ & $\delta_{2000}$ & $V$ & $B-V$ &
  $L$/\Lsun & \teff & $R$/\Rsun &
  $\delta L/L$ & $\nu_{\rm max}$ & $\Delta\nu_{0}$  & $\omega_c$
   \\
\hline
\hline                                   %inserts single line
   66 &   364   &   16 23 53.65  & -26 33 55.0 &    12.93 &    1.263 &    143.0 &    4723  &     17.9 &     254  &      8.9 &     1.64 &     16.4\\
   78 &  1715   &   16 23 32.46  & -26 26 20.7 &    13.09 &    1.254 &    123.1 &    4741  &     16.5 &     227  &     10.5 &     1.86 &     19.4\\
   83 &   689   &   16 23 27.34  & -26 30 59.1 &    13.03 &    1.303 &    130.3 &    4637  &     17.7 &     247  &      9.2 &     1.67 &     16.9\\
   84 &   264   &   16 23 54.72  & -26 31 22.6 &    13.12 &    1.183 &    120.0 &    4903  &     15.2 &     209  &     12.2 &     2.10 &     22.3\\
  126 & NA &   16 23 26.94  & -26 30 43.9 &    13.28 &    1.267 &    104.1 &    4714  &     15.3 &     204  &     12.2 &     2.07 &     22.4\\
  162 &   986   &   16 23 50.23  & -26 29 28.0 &    13.41 &    1.181 &     92.4 &    4909  &     13.3 &     173  &     15.8 &     2.56 &     29.1\\
  166 &  1753   &   16 23 22.96  & -26 30 32.6 &    13.44 &    1.221 &     89.3 &    4815  &     13.6 &     176  &     15.3 &     2.48 &     28.2\\
  181 & NA &   16 23 50.01  & -26 32 21.5 &    13.46 &    1.209 &     88.0 &    4844  &     13.3 &     172  &     15.9 &     2.55 &     29.2\\
  195 & NA &   16 23 47.55  & -26 31 06.0 &    13.47 &    1.218 &     86.9 &    4822  &     13.4 &     172  &     15.8 &     2.54 &     29.1\\
  200 & NA &   16 23 19.94  & -26 32 40.1 &    13.54 &    1.203 &     82.0 &    4856  &     12.8 &     163  &     17.2 &     2.71 &     31.6\\
  201 & NA &   16 23 53.03  & -26 28 09.1 &    13.53 &    1.201 &     82.4 &    4861  &     12.8 &     163  &     17.2 &     2.71 &     31.6\\
  210 & NA &   16 23 43.11  & -26 28 08.8 &    13.52 &    1.190 &     83.4 &    4888  &     12.8 &     163  &     17.3 &     2.73 &     31.8\\
  228 & NA &   16 23 47.84  & -26 29 50.2 &    13.56 &    1.196 &     80.5 &    4873  &     12.6 &     160  &     17.7 &     2.78 &     32.6\\
  229 & NA &   16 23 45.01  & -26 33 58.3 &    13.57 &    1.227 &     79.4 &    4803  &     12.9 &     163  &     17.1 &     2.68 &     31.4\\
  243 & NA &   16 23 21.77  & -26 26 45.8 &    13.59 &    1.205 &     77.9 &    4851  &     12.5 &     158  &     18.0 &     2.81 &     33.1\\
  244 & NA &   16 23 24.57  & -26 31 10.6 &    13.64 &    1.226 &     74.6 &    4805  &     12.5 &     156  &     18.2 &     2.82 &     33.5\\
  259 &   243   &   16 23 42.43  & -26 33 18.6 &    13.75 &    1.177 &     67.6 &    4919  &     11.3 &     139  &     21.8 &     3.26 &     40.1\\
  266 & NA &   16 23 48.97  & -26 29 21.3 &    13.70 &    1.278 &     70.8 &    4691  &     12.8 &     158  &     17.7 &     2.73 &     32.4\\
  268 &   331   &   16 23 41.72  & -26 29 47.9 &    13.73 &    1.213 &     68.6 &    4834  &     11.8 &     145  &     20.2 &     3.06 &     37.2\\
  270 &   540   &   16 23 47.83  & -26 32 44.7 &    13.78 &    1.153 &     65.4 &    4976  &     10.9 &     133  &     23.5 &     3.45 &     43.1\\
  280 &   463   &   16 23 32.34  & -26 29 22.4 &    13.82 &    1.172 &     63.0 &    4931  &     10.9 &     131  &     23.6 &     3.46 &     43.4\\
  285 &   305   &   16 23 42.13  & -26 32 07.4 &    13.83 &    1.233 &     62.7 &    4788  &     11.5 &     139  &     21.4 &     3.18 &     39.3\\
  292 &  1005   &   16 23 26.05  & -26 30 39.6 &    14.03 &    1.172 &     51.9 &    4931  &      9.9 &     115  &     28.7 &     4.00 &     52.7\\
  293 &  1059   &   16 23 22.50  & -26 29 41.2 &    14.07 &    1.165 &     50.1 &    4947  &      9.7 &     111  &     30.0 &     4.15 &     55.2\\
\hline
\end{tabular}
\end{table*}

\subsection{Removing outliers and weighting the data\label{sec:weights}}

Due to the changes in observing conditions
there is a large variation in
the data quality from night to night. 
It is therefore essential to apply weights when the Fourier spectra are calculated. 

%% Weights are calculated separately for each star.

In the first step we remove obvious outliers. All points that
deviate more than 5$\sigma$ from the mean are discarded. 
Weights are then calculated for each star by combining 
a weight for each night and a weight based on the point-to-point scatter. 
Each night gets a weight, which is 
\begin{equation}
w_1 = 1/\sigma_{night}^2,
\end{equation}
where $\sigma_{night}$ is the \rms\ noise of the time series for the entire night.
All data points from this night get the same weight.
% {\bf DS: average} noise per data point {\bf DS:
% Weights are calculated separately for each star.
% what do you mean by noise per data point, rms of the time series of that
% night, or point-to-point scatter of time series or what, please be clear}
% around the mean for the whole night{\bf DS: it is not clear if the night
% weight is the same for all stars }. 

The point-to-point  weight 
% {\bf DS: do you mean point-to-point weight} 
is calculated by first computing a spline
fit to an averaged time series with a smoothing width of $\pm 0.075$\,d. 
The spline fit is subtracted from the data. 
% {\bf HB: give box-width of spline -- how many data points are averaged, eg.\ expressed as a time interval?}.
% {\bf DS+HB: We do not understand the next sentence. Is the weighted average
%   using the Gauss as weight function? Did you just calc the average of the
%   residuals? We would expect that to be about zero plus minus a bit, meaning
%   we would have negative values as well which does not make sense. 
%   Why not calc the average RMS of the residuals?}
The noise level at any given point in time $\sigma_{pt}$ is then
the average \rms\ calculated using a
Gaussian with a width of 0.05\,d. 
% {\bf HB: what is the FWHM of this Gaussian (in time units)?} 
% Outliers {\bf HB: what is the criterion: 4\,$\sigma$?} 
% are eliminated during this process{\bf DS: is this a second round of sigma
%   clipping after the 5$\sigma$ clipping mentioned a few paragraphs earlier,
%   please be clear on that and more specific about the clipping process?}.
% It was not a real clipping, rather clipping of odd points (dm = 99. and alike)
The weight is chosen as the inverse of the noise level:
\begin{equation}
w_2 = 1/\sigma_{pt}.
\label{eq_w2}
\end{equation}

We note that Eq.~\ref{eq_w2} (rather than use of inverse variance) was also the preferred 
weighting in si\-mi\-lar 
multi-site campaigns by Handler (\cite{handler03}) and Bruntt et al.\ (\cite{brunttM67}).

The final weight we use is the product of $w_1$ and $w_2$.
A maximum weight (4 times the mean weight)
% {\bf HB: specify this max.\ weight?}
is defined to avoid nights with very few points to give very high weights.

% {\bf HB: Show the effect of using the weights. Show how the noise is improved at
% high frequencies versus magnitude.}

\section{Selection of K~giant targets}
We selected two dozen K~giants for a detailed analysis.
The stars are found in the least crowded parts of M4 and cover a 
range in luminosity on both sides of the bump stars.
Further, we selected the stars with the \rms\ scatter in the time series 
less than 4 mmag.
We list their ID numbers, standard $V$ magnitude and $B-V$ colour in Table~\ref{table:3}.
The position of some of the stars in the colour-magnitude diagram 
is shown in Fig.~\ref{fig:1a} and the position
in the cluster are marked with squares in Fig.~\ref{fig:2}.

To calculate $L$/\Lsun, \teff\ and $R$/\Rsun\ for the targets we used cluster parameters from 
Ivans et al.\ (\cite{ivans}) for M4: distance $d = 2.1$~kpc, reddening $E(B-V) = 0.37$,
interstellar absorption $A_V = 4.0E(B-V)$, metal content [Fe/H]$ = -1.2$,
bolometric correction $BC = -0.48$ and mass $M = 0.85\Msun$.
As pointed out by Ivans et al.\ (\cite{ivans}), the reddening is varying over the field of the cluster.
These parameters lead to a distance modulus of $(m-M)_V=13.1$. 
This compares reasonably well with the mean magnitude of the RR~Lyrae 
stars (Kopacki \& Frandsen, \cite{rrlyr})
in the range $13.2$--$13.3$, which have absolute magnitude $M_{\rm V}\simeq0.0$. 
The uncertainty of the cluster parameters introduce a 
$\sim$10\% uncertainty in the luminosities.

\subsection{Expected signal for solar-like oscillations\label{sec:expect}}

We have made estimates of the amplitude and frequency that is expected 
for solar-like oscillations. 

We use the following scaling relations for the
frequency separation $\Delta \nu_0$, the expected 
location of $p$-mode maximum frequency $\nu_{\rm max}$ and the
acoustic cutoff frequency $\omega_c$:
\begin{equation}
\Delta \nu_0 = 134.9\,(M/\Msun)^{1/2}(R/\Rsun)^{-3/2}\,\mu{\rm Hz}, \label{eq:start}
\end{equation}
\begin{equation}
\nu_{\rm max} = 3050\,{M/\Msun \over (R/\Rsun)^2(T_{\rm eff}/5777\,{\rm K})^{0.5}}\,\mu{\rm Hz}, \label{eq:numax}
\end{equation}
\begin{equation}
\omega_c = 5600\,{(M/\Msun) \over (R/\Rsun)^2(T_{\rm eff}/5777\,{\rm K})}\,\mu{\rm Hz}.          \label{eq:acoustic}
\end{equation}
The expected amplitude of the oscillation modes 
were estimated from the calibration by Kjeldsen and Bedding (\cite{kandb}), but using the
scaling $\delta L/L\propto (L/M)^{0.7}$ suggested by Samadi et al.\ (\cite{samadi}):
\begin{equation}
(\delta L/L)_{\lambda} = 4.7\,{[(L/\Lsun)\,/ \,(M/\Msun)]^{0.7} \over (\lambda/550\,{\rm nm}) \,
  (T_{\rm eff}/5777\,{\rm K})^2}\,{\rm ppm}. \label{eq:slut}
\end{equation}
% {\bf DS: Put units as the last thing in all the equations...not in the
%   middle of it all...throughout the paper.}
The computed parameters using Eq.~\ref{eq:start}--\ref{eq:slut} 
for the 24 selected K~giants are listed in Table~\ref{table:3}.
As the luminosity decreases from about 150 to 50\,$L_\odot$
the amplitudes decreases from 300 to 100\,ppm, 
while the frequency of maximum power shifts from 9 to 30\,\mhz\ (periods of 31 to 9 hours).
The search for the signatures of the oscillations will be difficult, 
as we are in the region of the frequency spectrum that is affected 
by slow drifts originating from observing conditions 
(seeing, transparency and extinction) and
the instrument (eg.\ slight changes in the 
position of the stars on the CCD).

\section{Time series analysis}

\subsection{Fourier analysis}

\begin{figure*}
   \centering
   \includegraphics[width=180mm]{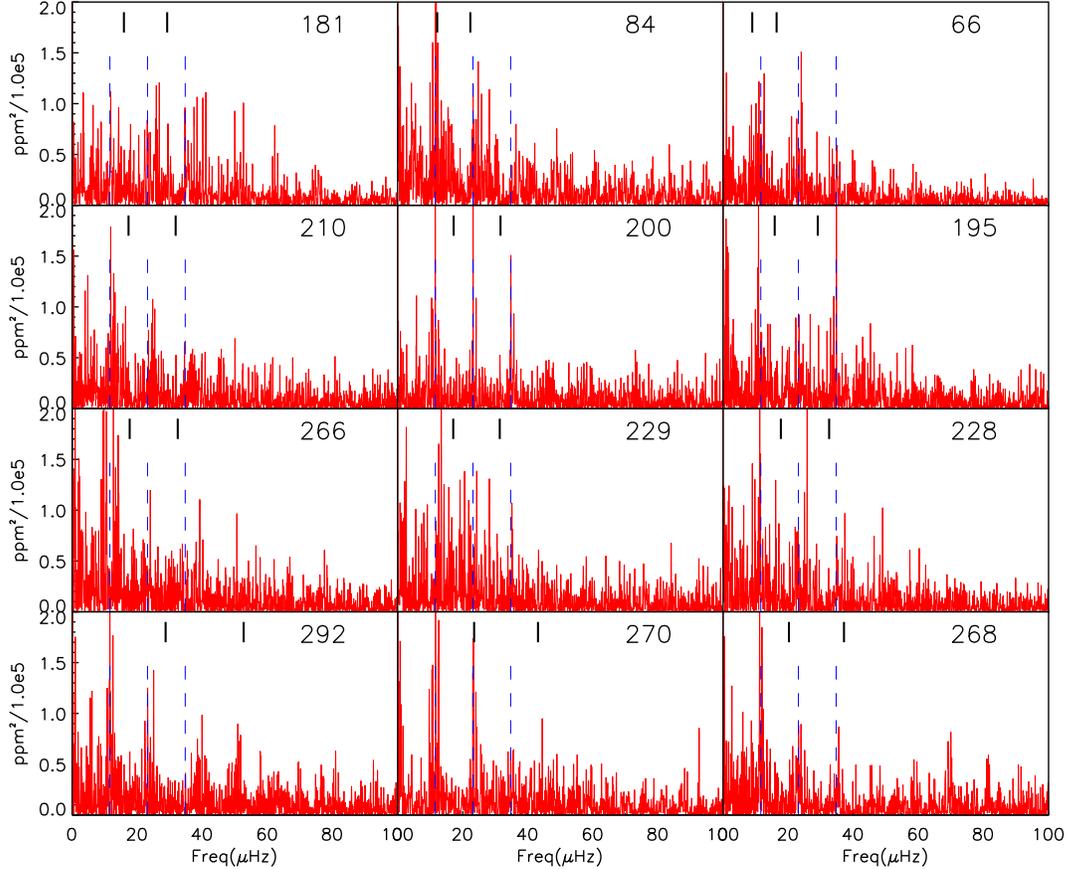}
   \caption{Power spectra for the 12 K~giants with the lowest noise at high frequencies. 
   The luminosity of the stars increases from the bottom to top panels
   and increases from left to right.
   The predicted location of the maximum $p$-mode power and the acoustic cutoff frequency
   (indicated by thick black lines) shift to lower frequencies for increasing luminosity.
   Frequencies at multiples of 11.57\,$\mu$Hz (1\,c/d) are indicated with vertical dashed lines.
              \label{fig:10a}}
    \end{figure*}

The light curves were analysed using the Fourier analysis program 
\period\ (Sperl \cite{sperl98}).
In Fig.~\ref{fig:10a} we show the
power spectra computed from the light curves using weights (\cf\ Sect~\ref{sec:weights}).
% {\bf DS: Only 12 are shown, not 24 as said in text.}
The spectra of a dozen K~giants covering a range in luminosity 
increasing from 50 to 140\,\Lsun\ from the bottom to the top panels.

In general the white noise increases as the stars become fainter but
some stars have significantly higher noise than stars with similar brightness, 
which in some cases is due to the presence of a close neighbouring star
but in other cases may be an unidentified instrumental problem.
The noise level in the amplitude spectra lies in the range 60--100\,ppm
at 81--104\,$\mu$Hz (white noise) increasing to 100--140\,ppm at 23--46\,$\mu$Hz.
With this low noise level we might be able to detect
oscillations if the amplitudes
follow the extrapolations used in Sect.~\ref{sec:expect} 
(\cf\ Table~\ref{table:3}) and the mode lifetime is not as short as a few days as found
by Stello et al. (\cite{lifetime}).
The stochastic excitation often gives rise to higher than
average amplitudes for observing periods that do not exceed
the lifetimes by large factors.

The observed amplitudes of the
highest peaks below the acoustic cutoff frequency $\omega_c$ range from 
200 to 400\,ppm in the spectra plotted in Fig.~\ref{fig:10a}. We cannot claim
that they are real as the signal-to-noise is low ($S/N < 4$) in all cases.
In a few stars we see suggestive evidence of power in the right range (\#181, \#228 and \#229),
which occurs between multiples of 1 c/d, but due to the low S/N
we cannot claim to have detected $p$-modes.
%% HB-urgent: paragraph above: the last 2 sentences in the preceding paragraph are not at all
%% clear to Dennis or Hans. Rephrase or say something more ``general'', e.g.:
%% In a few stars there may be indications of excess power, but due to the low
%% S/N we cannot claim to have detected ...
%%
%% SF - I have simplified it

\subsection{Quantifying the effect of the spectral window\label{sec:specwin}}

The observing window will introduce frequency spacings which
may be mistaken for the frequency spacing characteristic for solar
oscillations. The spectral window is different from star to star 
as the applied weights vary from star to star. 

\begin{figure}
   \centering
   \includegraphics[width=80mm]{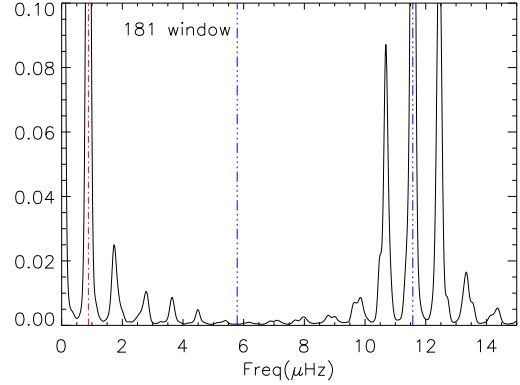}
   \caption{
      The autocorrelation of the spectral window for star
      \#181.  
      The vertical lines indicate peaks at 11.57, 5.84 and 0.81\,$\mu$Hz 
      (1, 0.5 and 0.07\, c/d). Note that multiples of the 0.81\,$\mu$Hz
      peak are present.
      }
              \label{fig:11}%
    \end{figure}

%% HB-June-Rev.: next paragraph rewritten slightly...
%%
In Fig.~\ref{fig:11} we show the autocorrelation function of 
the window function for star \#181. 
This is a representative example for the K~giant stars and
gives an idea of the effect of the complicated 
spectral window which is due to gaps in the time series.
The autocorrelation function has a very pronounced peak
at 11.57 $\mu$Hz due to the daily side lobes. In addition a
strong peak at $\sim$0.81$\mu$Hz is seen.
The origin of the peak is found in the observing schedule, 
where dark time was not allocated to the project (Fig.~\ref{fig:1}). 
When searching for the presence of the large separation using
the autocorrelation technique this will give rise to extra peaks.
For a few stars the window function shows a more complex 
pattern giving rise to even more peaks in the autocorrelation.
This will considerably reduce our ability to see clear evidence of
stellar signal oscillations (\cf\ Sect.~\ref{sec:auto}).

%%%
%%% HB-urgent: Fig. below: It is confusing that the luminosity is increasing from RIGHT to LEFT.
%%% Both Dennis and I thought so. Simply swap 66 and 181, 195 and 210, 228 and 266,
%%% and finally 268 and 292. Thanks.
%%%
%%% SF - IT IS INCREASING FROM LEFT TO RIGHT, no change is needed
%%% Lower numbers 

\subsection{Subtracting 1~c/d aliases}

We note that some of the observed excess peaks in the amplitude spectra 
are close to multiples of 1 c/d (11.57 \mhz)
as indicated by the vertical dashed lines in Fig.~\ref{fig:10a}.
These peaks are likely due to residual effects of extinction.
In some light curves a peak in the range $\nu$ = 0.8--0.9\,$\mu$Hz was 
also detected as likely caused by the spectral window (\cf\ Sect.~\ref{sec:specwin}).
Therefore, following the approach of Stello et al. (\cite{stello06}), 
we subtract these low frequency peaks if they are significant ($S/N>4$).
We specify later in each case, whether we are using the raw or the cleaned
spectra (\cf\ Sect.~\ref{sec:evidence}).

\begin{figure}
   \centering
   \includegraphics[width=80mm]{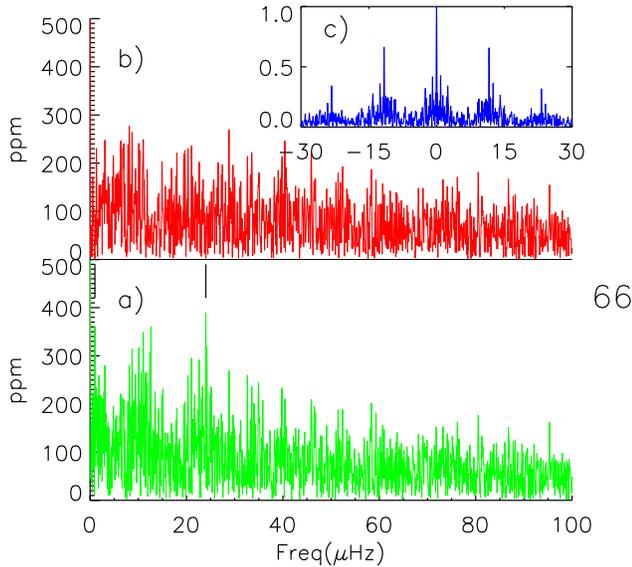}
   \caption{
       Amplitude spectra for star \#66 which is the brightest, non-saturated star
       in the subsample chosen for a detailed analysis. 
       Panel (a) is the raw spectrum with one low frequency 
       peak and a peak close to 2 c/d (23.14 \mhz) marked. 
       Panel (b) shows the amplitude spectrum after subtracting
       the two peaks marked with vertical lines in panel (a).
       Panel (c) shows the spectral window, which is seen to contain a 
       complex substructure around each 1 c/d sidelobe.
       }
              \label{fig:12}%
    \end{figure}

In Fig.~\ref{fig:12} we show an example of the process for star \#66.
Daily alias variations are seen in the amplitude spectrum in 
panel (a). Panel (b) is the amplitude spectrum after 
subtracting two peaks indicated by the vertical tickmarks in panel (a).
Panel (c) is the spectral window, which shows quite strong and complex sidelobes. 
The white noise level is $\simeq60$\,ppm and 
is calculated as the mean level in the frequency range 81--104$\,\mu$Hz.

\subsection{Light curve simulations\label{sec:sim}}

% {\bf Paragraph rewritten/expanded by HB+DS:}

To facilitate the interpretation of the observations in Sect.~\ref{sec:evidence}
we made simulations of the light curves. 
The time sampling and the noise properties are the same as for 
the observations and they include a known oscillation signal.

To mimic the noise we used the observed time series of two stars assuming that they resembled pure noise.
The two stars (\#66 and \#292) are at either end of the luminosity range of
the selected K~giants. Both are among the K~giant stars with
the lowest noise in the amplitude spectra. 
The two stars show no clear evidence of power excess or a frequency separation 
that could arise from stellar oscillations. 

We used different input amplitudes and life time of the modes using 
the simulation software
described by De~Ridder et al.\ (\cite{p-simul}) and Stello et al.\ (\cite{stello04}).

\section{A search for evidence of solar-like oscillations\label{sec:evidence}}

Inspection of the K~giant amplitude spectra in Fig.~\ref{fig:10a} 
reveals that the noise increases towards low frequencies, 
indicating that drift noise is present in the light curves.
This will seriously hamper the interpretation of the signal below $\simeq20$\,\mhz.
We see no clear evidence of excess power or the comb-like pattern 
expected for solar-like oscillations in any of the K~giants.
Although we find clear evidence for an increase in excess power 
towards low frequencies, it is not possible to disentangle drift 
noise and variation intrinsic to the K~giant stars,
eg.\ due to granulation or $p$-modes. 

We have used two techniques to look for evidence of $p$-modes in the amplitude spectra:

\begin{itemize}
\item Autocorrelation of the amplitude spectra to search for the large separation (Sect.~\ref{sec:auto}).
\item A search for evidence of excess power in the expected frequency range (Sect.~\ref{sec:excess}).
\end{itemize}

\subsection{Autocorrelation\label{sec:auto}}

\begin{figure}
   \centering
   \includegraphics[width=80mm]{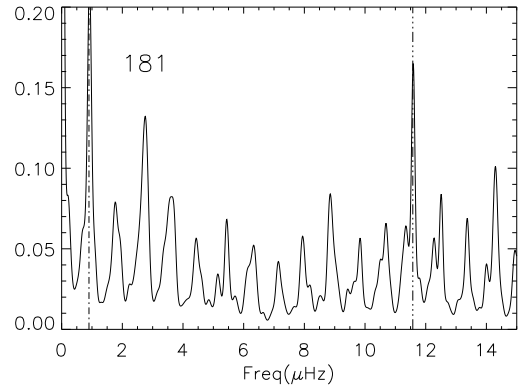}
   \caption{The autocorrelation of the raw power spectrum for 
      the giant star \#181. Vertical lines represent peaks in the window function.
      The peak at 2.75\,\mhz\ could be the large separation, but
      might also be due to low frequency noise interacting with the 
      window function.
      }
              \label{fig:12a}%
    \end{figure}

A clear proof of the presence of stellar oscillations would be a
systematic behaviour of the observed frequency separations. 
The separation should increase with decreasing luminosity (cf.\ Eq.~\ref{eq:start}). 
In a few cases the autocorrelation shows a peak that could
be indicative of the large separation. A prominent peak
is seen in star \#181 as shown in Fig.~\ref{fig:12a}, where there is
a peak at 2.75\,\mhz\ (predicted to be at 2.55\,\mhz). 
The highest peak at $\simeq$0.9\,\mhz\ (vertical dashed line)
is due to the window function (\cf\ Fig.~\ref{fig:11})
The autocorrelation was done on the raw spectrum.
%%%
%%% HB-urgent: paragraph above: you only mention the result for one star. 
%%% Perhaps quickly summarize the conclusions drawn for the other 23 stars?
%%%
%%% SF - added the following
We find a few other stars with peaks in the autocorrelation spectrum
that might come from the presence of a $p$-mode spectrum, but in
the majority of cases we are not able to find a single prominent peak.

We made several simulations of stars \#66 and \#292 with known input amplitude and 
lifetime as described in Sect.~\ref{sec:sim}. 
We used two methods to search for the known large separations of the inserted modes. 
The first method was a simple autocorrelation technique and
the second method cut up the spectrum in small pieces and added these
pieces (like when an Echelle diagram is created) and then searches
for peaks. The latter method is described by Christensen-Dalsgaard et al. (\cite{kepler}). 
Both methods gave quite similar results.
If the oscillations have a $Q$\footnote{Q is oscillation lifetime
divided by period of the mode as used by Stello et al.\,(\cite{qfactor})
and presented for a set of stars in their Fig. 4} value
above 200, the correct large frequency separation $\Delta\nu_0$ is recovered in both stars 
for amplitudes down to 400 ppm.

\begin{figure}
   \centering
   \includegraphics[width=70mm]{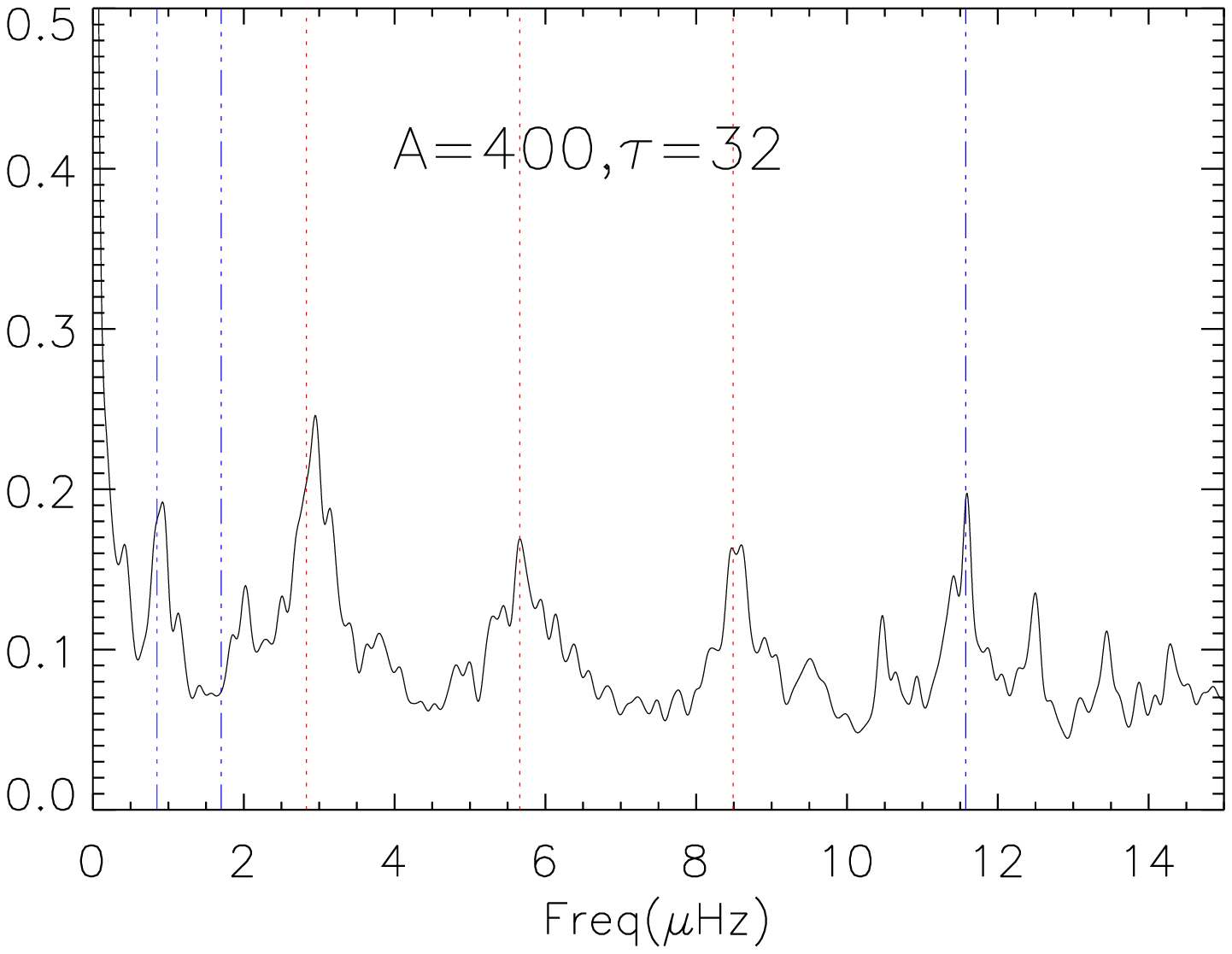}
   \includegraphics[width=70mm]{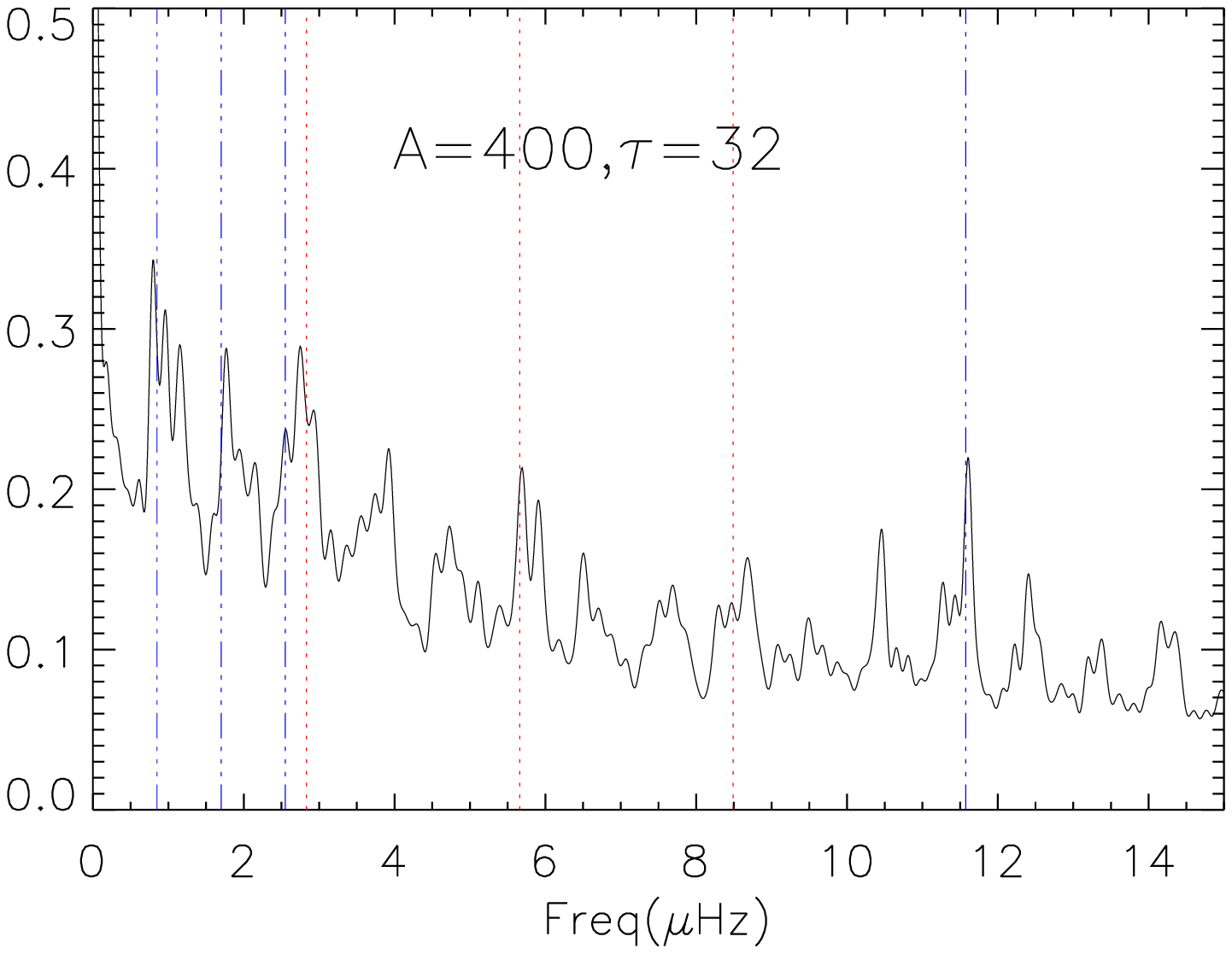}
   \caption{
   Autocorrelation of a simulation of star \#292 including oscillations.
   In the top panel the spectrum of \#292 was cleaned for multiples of 1 c/d before 
   adding the simulated signal and in 
   the bottom panel, the same procedure was followed as for the observed time series data.
   The red dotted lines correspond to multiples of the large separation in the simulated signal.
   The blue vertical lines indicate parasitic peaks
   from the window function at 0.85 and 11.57\,$\mu$Hz (see Fig.\,\ref{fig:12}).
   }
              \label{acor}%
\end{figure}

In Fig.~\ref{acor} we show the autocorrelation of the spectrum of star \#292 
including simulated oscillations with large separation 2.83\,$\mu$Hz, 
amplitude $A$=400\,ppm and lifetime $\tau$=32\,d. 
We note that this lifetime is at least a factor two longer than expected.
In the top panel the spectrum of \#292 was cleaned for multiples of 1 c/d before adding the simulated signal. 
The autocorrelation clearly shows several peaks  that correspond
to the large separation present in the simulated signal (red dotted lines).
The first dotted line indicates the large frequency separation
present in the simulation, which repeats several times.
The blue, dot-dash, vertical lines indicate {\em parasitic} peaks
in the window function at 0.85 and 11.57\,$\mu$Hz (see Fig.\,\ref{fig:11}).
In the bottom panel in Fig.~\ref{acor} the procedure applied
to the observed time series data was followed (see \#181 in Fig.\,\ref{fig:12a}).
In this case the input large separation is much harder to recover.

For the bright star (\#66) for inserted amplitudes as low as $A = 200$\,ppm 
and large $Q$ values ($>$30) the large separation is recovered. 
However, if the lifetime is shorter than $\simeq$16 days ($Q < 12$),  
%%% HB-urgent: please give the Q value when lifetime=16days !! The reader cannot do this off hand?
%%% SF - done!!
we find that it is not possible to recover the input separation in any of the
simulated series, even with amplitudes up to 800\,ppm.
Thus, our upper limits of 200--400\,ppm only hold if the lifetime
is greater than two weeks as predicted by Houdek \& Gough (\cite{houdek02})
and Houdek (\cite{houdek06}), and not a few days as 
found by Stello et al.\ (\cite{lifetime}).

%%% HB-urgent: original caption was VERY long. I moved it to the main text in Sec. 7.1 
%%% and rewrote the caption. Please chech that it is correct.

%%% HB-June-Rev.: Next section rephrased:

\subsection{A search for evidence of excess power\label{sec:excess}}

To search for excess power we followed the approach by 
Stello et al. (\cite{stello07}) used in their analysis of K~giants 
in the open cluster M67. The K~giant stars in M4 were put in three groups 
with luminosities around 60, 80, and 130 \Lsun\ (faint, average or bright stars).
For each group we calculated a mean raw amplitude spectrum.
To demonstrate the effect of cleaning the spectra, we also
produced a mean spectrum for the faint group of the cleaned spectra.

\begin{figure}
   \centering
   \includegraphics[width=80mm]{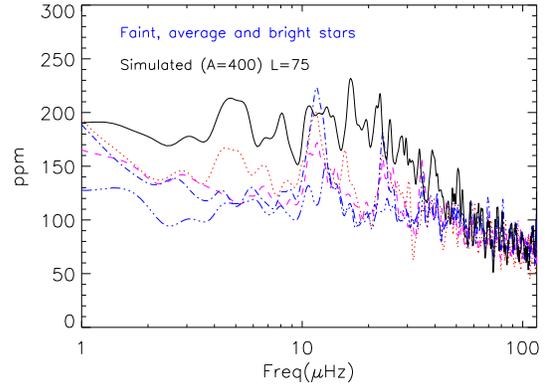}
   \caption{
   Average smoothed spectra for three groups of observed K~giants 
   with luminosity $L \simeq 60, 80 ~{\rm and}~ 130$\,\Lsun.
   The faint group (raw and clean version) are the blue lines, the average group is plotted
   as the magenta line and the bright group is the red line.
   The bottom curve is the faint group after multiples of 1 c/d are removed.
   The black solid line is for a simulation 
   of star \#292 with peak oscillation amplitudes of $A = 400$\,ppm.
   }
              \label{fig:comp1}%
\end{figure}

In Fig.~\ref{fig:comp1} we show the mean spectra.
The raw spectra for the three groups of stars are quite similar with
prominent peaks near multiples of 1 c/d.
This is unexpected and suggests that
the power seen possibly is dominated by a noise component due 
to instrumental drift in the data.

Using the simulations described in Sect.~\ref{sec:sim} we
probed our ability to identify the inserted oscillation signal.
The black, solid spectrum in  Fig.~\ref{fig:comp1} is based on a mean of four time series of 
star \#292 which includes simulations of $p$-modes using different lifetimes
in the range 2--8 days and peak amplitudes of $A=400$\,ppm per mode.
This covers the range of expected lifetimes and averaging four simulations 
removes some of the stochastic features of the oscillations.
This artificial spectrum is clearly above any of the observed mean spectra.
Our conclusion is that excess power from $p$-modes with amplitudes $A > 300$\,ppm 
would show up clearly in our observed spectra.

% At low frequencies the power spectrum of the flux of a star
% is dominated by granulation and stellar activity. 
% This is also done now for red giants (Collet et al. \cite{rg-simul}).
% In particular we used a model presented by Ludwig (\cite{ludwig}).

%%% HB-June-Rev.: Rephrased first two paragraphs:

\section{Comparison with 3D-hydrodynamical simulations of granulation\label{sec:ludwig}} % of granulation noise}

We have compared our results for the K~giants 
with 3D hydrodynamical simulations of granulation 
by Svensson \& Ludwig (\cite{g-simul}). 
The effect of granulation is modelled by simulations of the
convective motions in a box which is scaled to predict 
the effect expected to be observed when observing a spherical star.
The result of Svensson \& Ludwig (\cite{g-simul})
indicate that K~giants should have a low frequency 
component of granulation power similar to the noise levels
achieved for some of the K~giants in M4. 

\begin{figure}
   \centering
   \includegraphics[width=80mm]{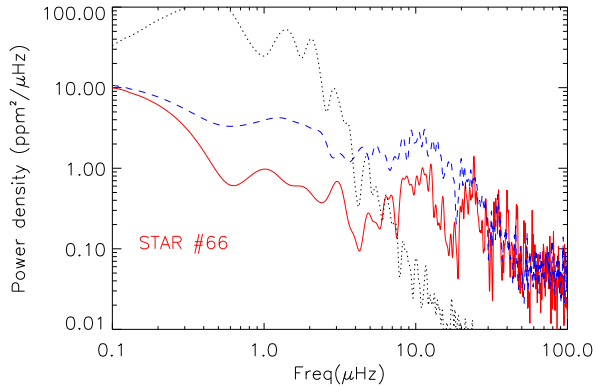}
   \caption{
   Power density for three cases: 1) the observed star \#66 (red, solid
   line), 
   2) a scaled power spectrum including granulation,
   $p$-modes with $A = 300$ppm and white noise as star \#66 (blue,
   dashed line), 3) the 
   simulation by Svensson and Ludwig (\cite{g-simul}) (black, dotted line).}
              \label{fig:gran}%
\end{figure}

One of their simulations (Ludwig \cite{ludwig}; kindly provided by G.\ Ludwig) 
is shown as the dotted black curve in Fig.~\ref{fig:gran}. 
The stellar parameters are $R/\Rsun = 30.2, \log g = 2.0,$ 
\teff\,$=4560$\,K, and metallicity [M/H]\,$=0.0$. 
This corresponds to a luminosity $L=385$\,\Lsun, which is
larger by roughly a factor two than the brightest K~giants we have observed.
The larger luminosity means that the simulated granulation
has larger amplitudes and excess power at lower frequencies than for our brightest target.

%%% HB-June-Rev.: next two paragraphs: please check content carefully, compare with old version!

A scaling of the solar power spectrum has been attempted
by Kjeldsen \& Bedding (in preparation) and by Stello et
al.\ (\cite{stello07}).
A white noise component with the same level as the star chosen for 
comparison (\#66) is added to a scaled granulation signal 
and a scaled $p$-mode spectrum. The scaling is very crude, 
and a difference of a factor two from our K~giants would not be surprising.
The granulation spectrum is shown in a logarithmic plot of power density in Fig.~\ref{fig:gran}.

The simulated spectrum (black dotted line) and the scaled spectrum (blue dashed line) both 
show a larger power density at low frequencies compared to the observations (red solid line).
The peak of the observed spectrum coincides with the
peak around 10\,$\mu$Hz of the scaled spectrum, in which the power is due
to the presence of oscillations. In the observed spectrum it is likely
to be the 1~c/d alias.

The reason for the high power at low frequencies in the 3D simulations is because
the luminosity of the model simulated is twice the luminosity of star \#66.
The amplitude of the power density should scale with $L^2$ and the frequency
with $L^{-1}$, which will bring the 3D simulations closer to the other
curves, but still above the observations.
Svensson and Ludwig (\cite{g-simul}) show that lowering the metal content in a
1~\Msun\ model makes
the power density drop considerably. A change of metal content might remove the discrepancy
between the observations and the 3D simulations of a giant.
The 3D simulation (black dotted line) falls below the other curves at high frequencies 
since no white noise was included.

\section{Conclusion and discussion}

We have presented results for a three-site multi-colour
photometric campaign on the globular cluster M4. 
We used data in the Johnson $BVR$ bands from 
48 nights with a time baseline of 78 days.

We compared the precision of the photometry obtained from 
three different photometric packages.
In the most crowded parts of the field we find the best results using a difference image analysis (Alard \cite{isis00}),
while classical aperture photometry give better results in the semi-crowded regions.
The best photometry is found for stars not affected by crowding, and in this case 
we achieve noise levels close to the theoretical noise limit.

%%%%
%%%% Paragraph below commented out by HB 3/5/2007: the conclusion should not have a lot of numbers
%%%% just PLAIN text explaining what we have found out.
%%%%
%%%%    In the frequency range 70--90\,$\mu$Hz we reach noise levels down to 60\,ppm 
%%%%    and the low frequency noise in the range 10--40\,$\mu$Hz is typically 100--140\,ppm.
%%%%    In the recent campaign on less luminous K~giants in the open cluster M67,
%%%%    Stello et al. (\cite{stello07}) obtained a better photometric precision at $\nu > 50\,\mu$Hz:
%%%%    they reached noise levels between 20--50\,ppm for several stars.
%%%%    To our knowledge our results are the best at $\nu < 50\,\mu$Hz obtained 
%%%%    from the ground for K~giants assuming the data by Stello et al. (\cite{stello07}) 
%%%%    are dominated by noise at low frequencies and not
%%%%    stellar signal (granulation and $p$-modes).
%%%%    This has only been achieved due to the careful differential 
%%%%    photometry possible in a dense cluster such as M4.

The main goal of the campaign was to detect solar-like oscillations in
the rich population of K~giant stars in M4. However, we 
are not able to claim an unambiguous detection and we summarize our conclusions here:

\begin{itemize}
\item
We detect no individual peaks with $S/N > 4.0$.
\item
We find no significant large frequency separation from the autocorrelation
of the amplitude spectrum. %% or other similar pattern searching techniques.
\item
We divided the stars with the best photometry in three groups and computed the average power spectra. 
From a comparison of the spectra with realistic simulations of granulation and an assumed comb-like
distribution of $p$-modes, we can exclude that power is present from
$p$-modes with peak amplitudes above 300~ppm whatever the damping time. 
This upper limit is in accordance with the $(L/M)^{0.7}$ scaling suggested by 
Samadi et al.\ (\cite{samadi}) which predicts amplitudes below 300\,ppm for the most luminous stars, 
but far below the scaling law $L/M$ from Kjeldsen \& Bedding (\cite{kandb}).
% Our results are in favour of the scaling from Samadi et al.\ (\cite{samadi}).
\item
In all K~giants the amplitude increases towards low frequencies. 
This would be consistent with granulation, but
we find that the observed increase does not vary with stellar luminosity.
Thus we cannot differentiate between drift noise in the data and the expected granulation signal.
\item
We have compared the observations with a 3D-hydrodynamical simulation for solar metallicity.
The observed power density at low frequencies is much lower than the simulations of granulations predict.
In their 3D simulations Svensson \& Ludwig \cite{g-simul} found that the granulation power 
is substantially smaller for lower metallicity (M4 has [Fe/H]\,$=-1.2$). 
This could explain that our data, even for the brightest star, falls considerably below the simulations
(Fig.~\ref{fig:gran}).
\end{itemize}

Our negative results support the recent evidence for short lifetimes 
in giant stars (Stello et al.\,\cite{lifetime}) but contradict the
theoretical predictions by Houdek \& Gough\,(\cite{houdek02}). 

For the population~II star $\nu$~Indi, which has a low luminosity (6\,\Lsun), 
$p$-modes have been detected (Bedding et al.\ \cite{bedding06}, Carrier et al.\ \cite{carrier07}). 
However, we cannot tell whether the stochastic
driving still works in population~II stars around $L=100$\,\Lsun,
although we can say it cannot be very efficient and certainly 
not exceeding amplitudes following the $(L/M)^{0.7}$ scaling relation 
suggested by Samadi et al.\ (\cite{samadi}). 

This is contrary to the conclusion by Stello et al. (\cite{stello07}) 
for the K~giants in M67, who found indications of $p$-mode power
in stars at lower luminosities (10--20\,\Lsun). 
They find better agreement between observations and predictions 
if the amplitudes scale as $L/M$. Their noise level at low frequencies do
not permit any statement for stars at luminosities similar to the
stars investigated in M4.
It remains an open question, whether the stochastic driving still
works without being destroyed by convective or radiative processes
in population~II K~giants. 

Based on the results obtained from large multisite campaigns 
on M67 and M4, it seems difficult to get conclusive positive results about
the stochastic oscillations and the granulation from ground-based
photometry. Improvements can be made, but extinction and other
atmospheric effects will always be a limiting factor and 
especially so at the time scales of the K~giants.
It is also difficult to get a clean window function due
to changing weather patterns for campaigns lasting several weeks, 
which is unfortunately needed to study solar-like oscillations in K~giant stars.

% One should keep in mind, that M4 has a metal content of [Fe/H]=-1.2, 
% whereas M67 has a solar composition.

% The low frequency increase in power is consistent with the expected
% granulation spectrum, but as the instrumental noise is of similar
% magnitude, we cannot separate the stellar and the instrumental 
% part of signal.

\subsection{Directions for the future\label{sec:future}}

If a new campaign is organized we recommend to use a single
filter, as colour information is not likely to be of a quality
that can lead to mode identification.
We found no use for the differences between the filters.
We recommend that the $V$ filter is used in the future:
the $B$ filter observations show strong colour-dependent 
extinction and the amplitudes will be very low in $R$.
%% HB-June-Rev.: I added the last two sentences above.

From the ground one would do better by organizing a longterm 
radial velocity campaign. One immediate advantage is that
the noise from granulation is 10 times lower in velocity (Grundahl et al. \cite{song}, their Fig. 1).
One could observe several K~giants during the night
since the time scale of the oscillations is several hours.
A dedicated network providing good time coverage would be
needed in order to obtain a clean window function.
The proposed SONG network 
(Grundahl, et al. \cite{song}) 
is an example of such a programme.

For fast rotating stars where the broad spectral lines prevent
precise velocity results to be obtained,
observations of solar-like oscillations can be made with photometry from space.
This has resulted in the measurements
of variability in population~I G~giants by the WIRE satellite
(Buzasi, Bedding \& Retter \cite{wire1}), 
and $p$~modes in $\eta$~Boo with the MOST satellite (Guenther et al.\ \cite{most}).

Giants are also among the targets for the new photometric satellite missions
like the CoRoT mission (Baglin, Michel \& Auvergne \cite{corot}) and
the future Kepler mission (Christensen-Dalsgaard et al. \cite{kepler}).
The CoRoT mission observes the same part of the sky for up to 150 days 
while the Kepler mission will cover many more targets for up to six years.
% {\bf HB: will any of the CoRoT stars be giants? Perhaps in the exo-planet field?}
% SF: Yes, there are giants like G8III, so I removed the letter K

\begin{acknowledgements}
This work was supported by the Danish National Research Council
and the Australian Research Council.
\end{acknowledgements}

\end{document}